\documentclass[10pt]{book}
\usepackage{array}

\usepackage{amsmath}
\usepackage{amsthm,graphicx}
\usepackage[ruled,vlined]{algorithm2e}
\usepackage{bbm}
\graphicspath{{}}
\usepackage[sectionbib]{natbib}
\usepackage{setspace}

\def\rji{r_{j,i}}
\def\pji{p_{j,i}}

\def\bX{{\boldsymbol X}}
\def\bR{{\boldsymbol R}}
\def\bP{{\boldsymbol P}}

\newtheorem{prop}{Proposition}[section]

\begin{document}


\renewcommand{\baselinestretch}{1.2}

\markboth{\hfill{\footnotesize\rm FLORE HARL\'E, FLORENT CHATELAIN, C\'EDRIC GOUY-PAILLER AND SOPHIE ACHARD} \hfill}
{\hfill {\footnotesize\rm BAYESIAN MODEL FOR MULTIPLE CHANGE-POINTS DETECTION} \hfill}
\renewcommand{\thefootnote}{}
$\ $\par


\fontsize{10.95}{14pt plus.8pt minus .6pt}\selectfont
\vspace{0.8pc}
\centerline{\large\bf BAYESIAN MODEL FOR MULTIPLE CHANGE-POINTS}
\vspace{2pt}
\centerline{\large\bf DETECTION IN MULTIVARIATE TIME SERIES}
\vspace{.4cm}
\centerline{F.~Harl\'e$^{\ast,\dagger,\ddagger}$, F.~Chatelain$^{\dagger,\ddagger}$, C.~Gouy-Pailler$^{\ast}$, S.~Achard$^{\dagger,\ddagger}$}
\vspace{.4cm}
\centerline{\it $^\ast$ CEA, LIST, Laboratoire d'Analyse de Donn\'ees et Intelligence des Syst\`emes,}
\centerline{\it 91191 Gif-sur-Yvette CEDEX, France}
\centerline{\it$^{\dagger}$ Univ. Grenoble Alpes, GIPSA-Lab,}
\centerline{\it F-38000 Grenoble, France}
\centerline{\it$^{\ddagger}$ CNRS, GIPSA-Lab,}
\centerline{\it F-38000 Grenoble, France}
\vspace{.55cm}
\fontsize{9}{11.5pt plus.8pt minus .6pt}\selectfont


\begin{quotation}
\noindent {\it Abstract:}
This paper addresses the issue of detecting change-points in multivariate time
series. The proposed approach differs from existing counterparts by
making only weak assumptions on both the change-points structure across series,
and the statistical signal distributions. Specifically change-points are not
assumed to occur at simultaneous time instants across series, and no specific
distribution is assumed on the individual signals.

\noindent It relies on the combination of a local robust statistical test
acting on individual time segments, with a global Bayesian framework able to
optimize configurations from multiple local statistics (from segments of a
unique time series or multiple time series).

\noindent Using an extensive experimental set-up, our algorithm is shown to
perform well on Gaussian data, with the same results in term of recall and
precision as classical approaches, such as the fused lasso and the Bernoulli
Gaussian model. Furthermore, it outperforms the reference models in the case
of non normal data with outliers. The control of the False Discovery Rate by an
acceptance level is confirmed. In the case of multivariate data, the
probabilities that simultaneous change-points are shared by some specific time
series are learned.

\noindent We finally illustrate our algorithm with real datasets from energy monitoring
and genomic. Segmentations are compared to state-of-the-art approaches based on
fused lasso and group fused lasso.
\par

\vspace{9pt}
\par
\end{quotation}\par


\fontsize{10.95}{14pt plus.8pt minus .6pt}\selectfont
\setstretch{1.2}

\setcounter{chapter}{0}
\refstepcounter{chapter}
\noindent {\bf 1. Introduction}
\setcounter{equation}{1}
\label{chap:intro}

\setcounter{section}{0}
\refstepcounter{section}
\noindent{\bf 1.1. Generalities}
\label{sec:Generalities}

The objective of change-points detection is to identify and localize abrupt changes in the time series.
This very active field of research has attracted lots of interest in various
topics where the time series can be modeled by homogeneous contiguous regions separated by abrupt changes.
These time series or signals are observed in different applications ranging from bioinformatics \citep{Allison2002,Bleakley2011,Muggeo2011}
to industrial monitoring \citep{Basseville1993}, but also, for example, applications in audiovisual data \citep{Desobry2005,Harchaoui2009,Kim2009} and
financial data \citep{Bai2003,Talih2005}. 
Our framework is the off-line joint estimation of time series.
This is a challenging problem in signal processing and statistics: in terms of data modeling, number of change-points to detect and size of the data. \\

Most of the methods developed for multiple change-point detection assume the time series to be piecewise Gaussian.
In the case of detecting one change-point, the Gaussian assumption allows to derive exact approaches based on maximum likelihood \citep{Hawkins2001}.
Other Bayesian methods have also been developed using a Bernoulli-Gaussian modeling \citep{DobigeonApril,Bazot2010}.
In the case of simple Gaussian models, Fearnhead \citep{Fearnhead2006} builds a Bayesian approach based on independent parameters
that avoids the need for simulations. However from a practical point of view, real datasets present often non Gaussian behavior,
and the classical approaches may fail. In order to propose a model-free approach, we choose to build our model based on an inference
function from the $p$-values of a statistical test computed on the data. The choice of the statistical test introduces a
crucial free parameter in the model, choosing a t-test or Welch's t-test is similar to the Bernoulli-Gaussian model developed
previously \citep{DobigeonApril}.
In this paper, we propose to use the Wilcoxon test so as to be free to the Gaussian assumption and to be robust to outliers. \\

When the objective is to identify multiple change-points, the computational cost is increasing drastically with
the number of change-points present in the time series. Using classical approach with maximum likelihood
induces a combinatorial cost for the optimization task and cannot be solved whenever the number of change-points is greater than two. Solutions have been developed in order to achieve the optimization task, for example,
dynamic programming \citep{Bai2003,YutFong2011}, approaches specific for lasso-type penalties \citep{Lavielle2006,Massart2007} or
Bayesian approaches \citep{Paquet2007}. Our approach presented in this paper is based on the derivation of a posterior
distribution using prior on change-points. The change-points are modeled using Bernoulli variables for the indicators.
The solution is then obtained by estimating the Maximum A Posteriori (MAP) with a Gibbs sampler strategy. \\

Nowadays, the datasets are usually recorded using several sensors. As many physical quantities, like electrical
power on devices, temperature, humidity, are measured by several sensors, events appear
simultaneously in some or every time series, depending on an implicit structure
between observations. Detecting change-points in multivariate data is a classical but major issue in
many fields. The literature in signal processing suggests different approaches for the multiple change-point
detection problem in multivariate data. Existing univariate methods
(as presented in \citep{Basseville1993,Harchaoui2013,Zou2007}) can be extended to the multivariate case.
In \citep{YutFong2011}, the authors present a statistics inspired by the well-known Wilcoxon/Mann-Whitney
test on multivariate time series to detect multiple change-points, but the events number is not clearly defined.
Moreover these events are supposed to occur simultaneously on every time series, assuming a fully connected
structure between all measured quantities. Research has been done also in genomics in order to identify
multiple change-points in multi-sample data \citep{nowak.2011.1,vert.2010.1}. The authors developed a
fused lasso latent feature model in order to identify part of the observations that are changing
knowing that samples share common information.
The dependencies between time series are taken into account in another approach, that consists in
using vague prior knowledge about data distribution and their relationships.
The authors of \citep{DobigeonApril} present a joint segmentation method based on Bayesian sampling strategy,
with various distributions depending on the physical properties of data.
Bayesian inference efficiency depends on prior choice, and is less accurate if data distribution is not correctly inferred.
Moreover in Gaussian models, algorithms appear to be highly sensitive to outliers.

Based on the change-points detection approach developed on the univariate case,
we combine a robust non-parametric test of time series to Bayesian inference with
priors on sensors relationships. This allows us to assess the segmentation
jointly and either to force given relationships between the time series or to learn these relationships.\\

\setcounter{section}{1}
\refstepcounter{section}
\noindent {\bf 1.2. Problem formulation}
\label{sec:pb_formulation}

Our goal here is to detect and localize the multiple change-points in time series.
We consider a signal of length $N$, stored in a vector $X$ where the sample $x_i$ denotes the observation at time $t$ such that $i < t \leq i+1$. The samples
$x_i$, $1\le i\le N$, are assumed to be mutually independent.
A change-point is defined as
the last point of a segment $s$ whose samples  $(x_i)_{i \in s}$  share some similar statistical properties. For instance, depending on the change hypotheses
these samples can have the same mean or median, be identically distributed according to a given or unknown distribution, ...

In order to model the presence or absence of the change-point at the different time instants, a vector $R \in \{0,1\}^N$
is introduced where an entry $r_i$ is an indicator variable such that
\begin{equation}
	r_i = \left\{
	\begin{array}{rl}
	1 & \text{if $x_i$ is a change-point},\\
	0 & \text{otherwise,}
    \end{array} \right.
\label{def R}
\end{equation}
for all $1\le i \le n$, with by convention $r_1=r_N=1$.

As a consequence, detecting the change-points is now equivalent to infer the indicator vector $R$. In a Bayesian framework, these estimates are
deduced from its posterior distribution $f(R|X)$ which classically expresses as
\begin{equation}
f(R|X) \propto L(X|R) f(R),
\label{bayes}
\end{equation}
where $f(R)$ denotes the prior on $R$ and $L(X|R)$ is the likelihood of the data.

In section~\ref{chap:uni_model}, we introduce our Bayesian change-point model, called the Bernoulli-Detector.
Extensive experimental set-up is given to quantify the performances of the proposed algorithm.
Then, in section~\ref{chap:multi}, the model is extended to the multivariate case.
Results of simulations are presented to empirically validate the model and to evaluate the performances,
with comparisons with other classical methods. Then two applications on real data are shown to illustrate the
different approaches of the multivariate model, in section~\ref{chap:appli_multi}. Finally the results are
discussed in section~\ref{chap:disc_concl}.
\\


\setcounter{chapter}{1}
\refstepcounter{chapter}
\setcounter{equation}{1}
\noindent {\bf 2. Bernoulli detector model}
\label{chap:uni_model}

This section is devoted to the introduction of the Bernoulli detector model. First, the observation model given a configuration of  change-points  is explained. It allows us to derive  in part~\ref{sec:likelihood}
a composite marginal likelihood in order to approximate the full likelihood $L(X|R)$ in \eqref{bayes}.
This data term is based on the $p$-values derived from any statistical test to detect a change.
To be free to the Gaussian assumption and to be robust to outliers, we will focus on the non-parametric Wilcoxon rank sum test recalled in part~\ref{sec:test_wilcoxon}.
The prior $f(R)$ in \eqref{bayes} chosen for the change-point indicator vector
is given in part~\ref{sec:prior_uni}, and the resulting posterior density is given in part~\ref{sec:posterior_uni}.
The algorithm is presented in part~\ref{sec:algo_uni}, and results on simulated data are given in part~\ref{sec:simu_uni}.
\\

\setcounter{section}{0}
\refstepcounter{section}
\noindent {\bf 2.1. Composite marginal likelihood}
\label{sec:likelihood}

So as to yield a surrogate of the likelihood function $L(X|R)$ in equation~\eqref{bayes}, the idea is to build an inference function from the
$p$-values of a statistical test, applied on each observation. Each $p$-value $p_i$ is considered as a random variable,
computed from the data. The test is applied on each $x_i$, for $2 \le i \le N-1$, given the indicators vector $R$.
Under the null hypothesis $H_0$ that $x_i$ is not a change-point, 
the $p$-value $p_i$ follows a uniform distribution over $[0,1]$, see for example~\citep{Sack99,Sellke01}.
Under the alternative hypothesis $H_1$, i.e. when $x_i$ is a change-point, this distribution is unknown and should depend on
the real distribution of the data.
However,  the $p$-values tend to be smaller than under $H_0$ for a consistent test.
Following an alternative proposition derived in~\citep{Sellke01} for the calibration of $p$-values,
we choose the class of Beta distributions $\mathcal{B}e(\gamma,1)$, with parameter $0 \leq \gamma \leq 1$.
Thus, the density of $p_i$ is decreasing, and the uniform distribution is the special case where $\gamma=1$.
It is interesting to note that this alternative Beta distribution leads to the following relationship
$F_1(p)= F_0(p)^{\gamma}$ where $F_0$ and $F_1$ are the  cumulative distribution functions (cdf)
under $H_0$ and $H_1$ respectively. This is a special case of Lehmann alternative \citep{Lehmann1975} to test
if a random variable is stochastically lower than the null hypothesis distribution. 
Here the parameter $\gamma \in (0,1)$ is expressed as a function of the acceptance level $\alpha$ such that 
\begin{align}
\label{eq:BetaParGen}
f( \alpha | r=1 ) = f( \alpha | r=0 ),
\end{align}
where $f( \cdot | r=1 )$ and $f( \cdot | r=0 )$ are the $p$-value densities under $H_1$ and $H_0$ respectively.
With the beta model for the alternative $H_1$, $\gamma$ is therefore the unique solution in $(0,1)$ of
\begin{align}
\label{eq:BetaPar}
\gamma \alpha^{\gamma-1} = 1,
\end{align}
for all $\alpha$ in $(0,e^{-1})$.
Note that if $\alpha \ge  e^{-1}$, then the unique solution of \eqref{eq:BetaPar} is $\gamma=1$. In this case,
the alternative distribution $H_1$ reduces to the uniform distribution $H_0$, and the model becomes inconsistent to identify
the null hypotheses and the alternative ones.
In the following, the parameter $\alpha$ is assumed to be in $(0,e^{-1})$.
Note that an interpretation of this acceptance level is given in section~\ref{sec:posterior_uni} 
for the single change-point detection.

This Beta-Uniform mixture modeling of the $p$-value distribution yields the following marginal density for each $p$-value:
\begin{align}
\label{eq:densP}
f( p_i | R ) &= \begin{cases}
                     \mathbbm{1}_{[0,1]}(p_i)  & \textrm{ if } r_i= 0 \textrm{ ($H_0$, $x_i$ is not a change-point)},\\
                     \gamma p_i^{\gamma-1} \mathbbm{1}_{[0,1]}(p_i) & \textrm{ if } r_i= 1 \textrm{ ($H_1$, $x_i$ is a change-point),}
                  \end{cases}
\end{align}
for all $2 \le i \le N-1$, where $\mathbbm{1}_{[0,1]}(\cdot)$ is the indicator function on $[0,1]$
($\mathbbm{1}_{[0,1]}(p)= 1$ if $p \in [0,1]$, $0$ otherwise).
\par
Finally, the data term of the model is formed by the following inference function:
\begin{align}
L_{\ast}(X|R)& = \prod_{i=2}^{N-1} f ( p_i | R )
 = \prod_{2=1}^{N-1} \left( \gamma p_i^{\gamma-1} \right)^{r_i}.
\label{eq:L_uni}
\end{align}
The term $L_{\ast}(X|R)$ in the expression~\eqref{eq:L_uni} is not a proper likelihood,
as defined in~\citep{Monahan1992}.  In fact, this data term is composed of a product
of the marginal likelihood of the $(p_i)_{1 \leq i \leq N}$ based
on their univariate  distributions and is termed a composite marginal likelihood \citep{varin2008,Varin2011}.
As a consequence, the dependencies between the $p$-values are not taken into account.
Then the coverage probabilities of posteriors sets should differ from the real ones.
The use of composite likelihood in a Bayesian framework has received recently some attention.
In \citep{pauli2011,ribatet2012}, the authors developed procedures based on moment matching conditions for
 a suitable calibration of the composite likelihood. This yields an
adjusted asymptotic for the posterior probability distribution. These methods are derived
for composite likelihood that are functions of some few continuous parameters, and
rely on the asymptotic normality of the maximum composite likelihood estimates.
Unfortunately, these assumptions are not in agreement with our model, where the parameters correspond to
the binary change-point indicators of the vector $R\in\{0,1\}^N$.
However, as explained in section~\ref{sec:posterior_uni} and \ref{ssec:FDR}, the acceptance level $\alpha$ that parametrizes the Beta distribution in \eqref{eq:densP}
yields to a weak local control on the change-point detection and a global control evaluated on simulations. Thus it induces a natural calibration
of the posterior distribution and can be used to control the detection performances.
\\

\setcounter{section}{1}
\refstepcounter{section}
\noindent {\bf 2.2. Wilcoxon test}
\label{sec:test_wilcoxon}

In order to be fully model-free,  we choose a non-parametric statistical test for the change-point detection.
It is important to note that this is a major advantage over classical Bayesian models, 
such as the Bernoulli-Gaussian model, especially when the $x_i$ cannot be assumed to be normally distributed. 
Furthermore, we choose a robust test based on the computation of ranks, namely the well-known Wilcoxon rank sum 
(\textit{aka} Mann-Whitney) test introduced in~\citep{wilcoxon1945}.

In our change-point detection framework,
the rule is to reject the null hypothesis that two populations defined by the observations belonging 
to two contiguous segments have the same median. The test statistic is computed as
\begin{equation}
U = \min(U_Y,U_Z)
\end{equation}
with
\begin{equation}
U_Y = n_Y n_Z + \frac{n_Y(n_Y+1)}{2}-R_Y,
\end{equation}
\begin{equation}
U_Z = n_Y n_Z + \frac{n_Z(n_Z+1)}{2}-R_Z,
\end{equation}
and where $n_Y$ and $n_Z$ are the sample size of the two segments denoted as $Y$ and $Z$ respectively, and
$R_Y$ and $R_Z$ are the sum of the ranks of the observations in $Y$ and $Z$ respectively, these ranks being the positions of the observations in the sorted vector $(Y,Z)$.
For small values of $n_Y$ and $n_Z$, the resulting $p$-values are tabulated from the exact distribution.
For larger samples, the $p$-values can be deduced from the asymptotic normal distribution of $U$. In fact, a standardized value can be computed as
\begin{equation}
z = \frac{U - m_U}{\sigma_U}
\end{equation}
where $m_U = \frac{n_Y n_Z}{2}$ and $\sigma_U = \sqrt{\frac{n_Y n_Z(n_Y+n_Z+1)}{12}}$. 

Finally, given the change-point indicator vector $R$ defined in \eqref{eq:densP}, each $p$-value $p_i$ for a possible change-point $x_i$, 
for $2\le i\le N-1$, is derived from the statistical test applied on the previous and next segments  $s_i^- = (x_k)_{i^-+1 \leq k \leq i}$ and $s_i^- = (x_k)_{i+1 \leq k \leq i^+}$ 
respectively, where $i^-$ and $i^+$ are the indexes of the previous and the next change-points respectively around the $i$th position in
$R$. It allows us to compute the inference function~\eqref{eq:L_uni}.
\\

\setcounter{section}{2}
\refstepcounter{section}
\noindent {\bf 2.3. Prior on indicators}
\label{sec:prior_uni}

The distribution of the indicators in $R$ in the equation~\eqref{bayes} is expressed given an hyperparameter $q$ 
that depicts the probability to have a change-point. For the observation $x_i$:
\begin{equation}
f(r_i|q) = q^{r_i} (1-q)^{(1-r_i)}
\end{equation}
The indicators $(r_i)_{2 \le i \le N-1}$ are assumed to be \textit{a priori} mutually independent, then:
\begin{equation}
f(R|q) = \prod_{i=2}^{N-1} q^{r_i} (1-q)^{(1-r_i)}.
\label{eq:f(R|q)}
\end{equation}
At last, in a hierarchical framework, the hyperparameter $q$ is also considered as a random variable. 
A Jeffreys noninformative prior is chosen for this hyperparameter, which leads to consider a Beta 
distribution $\mathcal{B}e(\frac{1}{2},\frac{1}{2})$.
\\

\setcounter{section}{3}
\refstepcounter{section}
\noindent {\bf 2.4. Posterior distribution}
\label{sec:posterior_uni}

The posterior distribution of the change-points indicators $R$ and the hyperparameter $q$ is derived 
from the relations~\eqref{eq:L_uni}, \eqref{eq:f(R|q)} and the prior  for $q$. It reads:
\begin{align}
f(R,q|X)& \propto  L_{\ast}(X|R)  f(R|q) f(q), \\
&\propto \left( \prod_{i=2}^{N-1}  (\gamma p_i^{\gamma-1})^{r_i} \right) \left( \Big[\prod_{i=2}^{N-1} q^{r_i} (1-q)^{1-r_i} \Big] \frac{1}{\pi} q^{-\frac{1}{2}}(1-q)^{-\frac{1}{2}}\right).
\label{eq:postR,q|X}
\end{align}
The conditional probability to have a change-point in $x_i$ is then deduced as
{\small
\begin{equation}
\begin{split}
&\Pr(r_i= \epsilon|r_2,\ldots,r_{i-1},r_{i+1},\ldots,r_{N-1},  q, X) =\\
&\frac{f(r_2,\ldots,r_{i-1},\epsilon,r_{i+1},\ldots,r_{N-1},q|X)}
{f(r_2,\ldots,r_{i-1},0,r_{i+1},\ldots,r_{N-1},q|X) + f(r_2,\ldots,r_{i-1},1,r_{i+1},\ldots,r_{N-1},q|X)},
\label{eq:condRi|X}
\end{split}
\end{equation}
}
for all $\epsilon\in \{0,1\}$. 

Based on this posterior, it is now possible to derive some properties of the
classical Bayesian estimates of a change-point for a given location $i$ under some simple hypotheses. We assume in the following propositions
that there is no other change-point in the signal. 
We denote as $R_0^{\backslash i}$ the void configuration event such that  $r_k=0$ for all $k \ne i$, and 
$q$ stands for the {\it  a priori} change-point probability $\Pr(r_i=1)=q$.

\begin{prop}{MAP estimator $\widehat{r}_i$ given $R_0^{\backslash i}$ and $q$. }
   Under the previous hypothesis, the conditional MAP estimate is
   \begin{align*}
      \widehat{r}_i \equiv \arg\max_{\epsilon} \Pr(r_i=\epsilon |R_0^{\backslash i},q,X)  &  = \begin{cases}
                 1 & \textrm{ if } \gamma p_i^{\gamma-1}  > \frac{1-q}{q}\\
                 0 & \textrm{ otherwise.}
                \end{cases}
   \end{align*}
   As a consequence, if $q = 1-q= 1/2$, then $\widehat{r}_i= 1$ iff the $p$-value is 
   lower than the chosen significance level $\alpha$,
   i.e. $p_i< \alpha$ according to \eqref{eq:BetaPar} and \eqref{eq:densP}.
   \label{prop:MAP}
\end{prop}

\begin{prop}{MMSE estimator $\widehat{r}_i$ given $R_0^{\backslash i}$ and $q$. }
    Under the previous hypothesis, the conditional MMSE estimate is
   \begin{align*}
    \widehat{r}_i \equiv E[r_i|R_0^{\backslash i},q,X] & = \frac{   \gamma p_i^{\gamma-1}  q  }{ 1-q +
     \gamma p_i^{\gamma-1} q }.
   \end{align*}
  If $q = 1-q = 1/2$, $\widehat{r}_i > 1/2$
  iff the $p$-value is lower than the chosen significance level $\alpha$, i.e. $p_i< \alpha$.
  \label{prop:MMSE}
\end{prop}
The proofs of these propositions are directly derived from the definition of these estimators, the expression of the 
posterior distribution \eqref{eq:condRi|X} and the choice of the parameter $\gamma$ in \eqref{eq:BetaPar}.

These properties illustrate the influence of the significance level $\alpha$ chosen in \eqref{eq:BetaPar}
to calibrate the distribution \eqref{eq:densP} under the alternative hypothesis for the single change-point problem. 
If the priors  on the configurations are equivalent then the presence of a change-point is favored when the support
against the null hypothesis is significant for the level $\alpha$.
\\

Finally, the hyperparameter $q$ can be viewed as a nuisance parameter and is now marginalized out: 
in \eqref{eq:postR,q|X}, one can see that the posterior of $q$ reduces to a beta distribution 
$\mathcal{B}e(K+\frac{1}{2},N-K-\frac{3}{2})$, where $K=\sum_{i=2}^{N-1}r_i$ is the total number of change-points.
After integrating $q$ in equation~\eqref{eq:postR,q|X}, the marginalized posterior expressed as follow:
\begin{align}
f(R|X) \propto \Gamma\left(K+\frac{1}{2}\right) \Gamma\left(N-K-\frac{3}{2}\right)  \prod_{i=2}^{N-1}  (\gamma p_i^{\gamma-1})^{r_i} .
\label{eq:postR|X}
\end{align}

\setcounter{section}{4}
\refstepcounter{section}
\setcounter{algocf}{0}
\noindent {\bf 2.5. Algorithm}
\label{sec:algo_uni}

To estimate the Maximum A Posteriori (MAP) of the posterior density~\eqref{eq:postR|X}, 
a Monte Carlo Markov Chain Method is applied, with a Gibbs sampler strategy. 
The best sampled change-point configuration, i.e. the configuration that maximizes the posterior distribution \eqref{eq:postR|X} 
among all the sampled configurations, is retained as the MAP approximation.

The pseudo-code of the implemented Gibbs sampler is given in algorithm~\ref{algo_uni_block}. 
At each MCMC iteration, the indicators are sampled for the time points $2 \leq i \leq N-1$, that are picked up randomly.
To sample the indicator $r_i$, the $p$-value $p_i$ is computed, but also 
the $p$-values $p_{i^-}$ and $p_{i^+}$ for the previous and the next change-points which have to be updated 
according to the new segmentation induced by the possible values of $r_{i}$.
To express the conditional distribution for the new indicators, we introduce the following notations: $R_{i-}$ denotes the vector of the indicators before $r_i$ (excluded), $R_{i+}$ denotes the vector of the indicators after $r_i$ (excluded), $R^{(m-1)}$ denotes vector of the indicators that have been updated at iteration $m-1$ and $R^{(m)}$ denotes vector of the indicators that have already been updated at iteration $m$.
Then, at iteration $m$ and time point $i$, $r_i^{(m)}$ is drawn from its conditional distribution

{\small
\begin{align*}
 \Pr( & r_i=\epsilon | R_{i-}^{(m-1)},R_{i-}^{(m)},R_{i+}^{(m-1)},R_{i+}^{(m)};X) =\\
 &
 \frac{ f ( \epsilon |R_{i-}^{(m-1)},R_{i-}^{(m)},R_{i+}^{(m-1)},R_{i+}^{(m)};X)  }
 {  f ( 0 |R_{i-}^{(m-1)},R_{i-}^{(m)},R_{i+}^{(m-1)},R_{i+}^{(m)};X) + f ( 1 |R_{i-}^{(m-1)},R_{i-}^{(m)},R_{i+}^{(m-1)},R_{i+}^{(m)};X) }.
\end{align*}
}

In addition, in order to improve the mixing properties of the sampler, 
a blocked Gibbs sampler is applied around the current change-points: if $r_i^{(m-1)}=1$, then at iteration $m$, 
the indicators $r_{i-1}$, $r_i$ and $r_{i+1}$ are sampled together from their joint conditional probability. 
\begin{algorithm}
\caption{Univariate Bernoulli-Detector, blocked Gibbs sampler}
\label{algo_uni_block}
\SetKw{Req}{\textbf{require}} \Req{$\alpha$}\\
\SetKw{Ini}{\textbf{initialize}} \Ini{$R^{(0)}$, $M$}\\
\For{$m\leftarrow 1$ \KwTo $M$}{
	initialize the index set $I = \{2,\ldots,N-1\}$ \\
	\While{$I \neq \emptyset $}{
		pick randomly $i$ in $I$ \\
		\If{$r_i^{(m-1)}=1$}{
				sample $r_{i-1}^{(m)}$, $r_{i}^{(m)}$ and $r_{i+1}^{(m)}$ from their joint conditional probability\\
				remove $i-1$, $i$, $i+1$ from $I$
				}
		\Else{
		sample $r_{i}^{(m)}$ from its conditional probability\\
		remove $i$ from $I$
		}
		}
	}
\Return{$R$}\\
\end{algorithm}

To reduce the computational cost of this algorithm, an approximation is done: 
at each iteration $m$ and time point $i$, only the $p$-value $p_i^{(m)}$ of the currently sampled $r_i^{(m)}$ is computed,
without considering the impact of a new segmentation on the $p$-values $p_i^-$ and $p_i^+$ 
of the previous and next change-points. 
When these $p$-values are not updated, the conditional change-point probability becomes:
\begin{align}
  &\Pr( r_i= 1 | p_i^{(m)}, K^{\backslash i} )  = 
  & \frac{ (K^{\backslash i} +1/2 ) \gamma \left(p_i^{(m)}\right)^{\gamma-1}}
  { (K^{\backslash i} +1/2 ) \gamma \left(p_i^{(m)}\right)^{\gamma-1} + N-K^{\backslash i} - 5/2 },
\label{eq:pseudoProp}
\end{align}
where $K^{\backslash i}$ is the number of change-points but $i$ in the current indicator vector. 
One consequence of this sampling procedure is that the mixing properties of the sampler are greatly improved, thus the blocked sampling approach is no longer necessary. 
This is a computational advantage for the analysis of multivariate data, as seen further in part~\ref{chap:multi}. 
In that case, the conditional probabilities does not form a compatible joint model and are said to be incompatible. 
A discussion about these potentially incompatible conditional-specified distributions (PICSD) is done in~\citep{Chena}. 
However the approximation is empirically justified in the section~\ref{sec:simu_uni} as it provides similar segmentation 
performances than the block Gibbs sampler. 
The new algorithm, with this pseudo-Gibbs sampling strategy, is described in algorithm~\ref{algo_uni_pseudo}. 
Performances of both algorithms are tested on simulated data.
\begin{algorithm}
\caption{Univariate Bernoulli-Detector, pseudo-Gibbs sampler}
\label{algo_uni_pseudo}
\SetKw{Req}{\textbf{require}} \Req{$\alpha$}\\
\SetKw{Ini}{\textbf{initialize}} \Ini{$R^{(0)}$, $M$}\\
\For{$m\leftarrow 1$ \KwTo $M$}{
	initialize the index set $I = \{2,\ldots,N-1\}$ \\
	\While{$I \neq \emptyset $}{
		pick randomly $i$ in $I$ \\
		compute $p_i^{(m)}$\\
		sample $r_i^{(m)}$ according to \eqref{eq:pseudoProp}  \\
		remove $i$ from $I$\\
		}
	}
\Return{$R$}\\
\end{algorithm}

\setcounter{section}{5}
\refstepcounter{section}
\setcounter{equation}{5}
\noindent {\bf 2.6. Simulations}
\label{sec:simu_uni}

\setcounter{subsection}{0}
\refstepcounter{subsection}
\noindent {\bf 2.6.1. Empirical performances for single change-point detection}
\label{ssec:single_CP_uni}

As a first validation, a signal of $N=100$ time points with a single change-point at $t=50$ is simulated. The data are generated following a normal distribution $\mathcal{N}(\mu_k,\sigma)$ for the $k$th segment, with different levels of noise. A signal-to-noise ratio (SNR) between to successive segments $k$ and $l$ is defined as follow:
\begin{equation}
SNR = 10\log \frac{(\mu_k-\mu_l) ^2}{\sigma^2}.
\label{eq:SNR}
\end{equation}

\begin{figure}[htb]
\centering
\begin{minipage}[b]{0.32\linewidth}
  \centering
  \centerline{\includegraphics[width=1.0\linewidth]{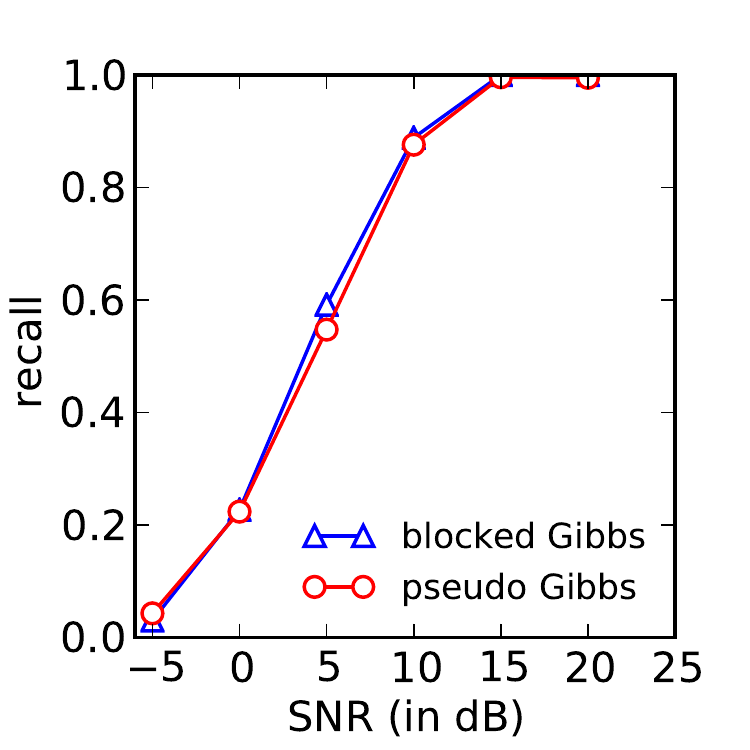}}
  \centerline{(a) Recall, $t=0$}\medskip
\end{minipage}
\begin{minipage}[b]{0.32\linewidth}
  \centering
  \centerline{\includegraphics[width=1.0\linewidth]{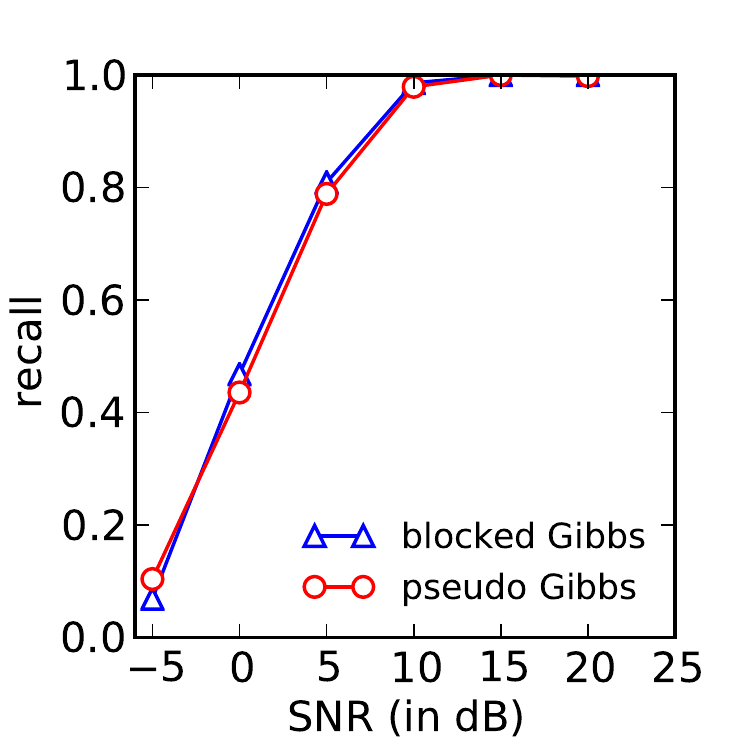}}
  \centerline{(b) Recall, $t=1$}\medskip
\end{minipage}
\begin{minipage}[b]{0.32\linewidth}
  \centering
  \centerline{\includegraphics[width=1.0\linewidth]{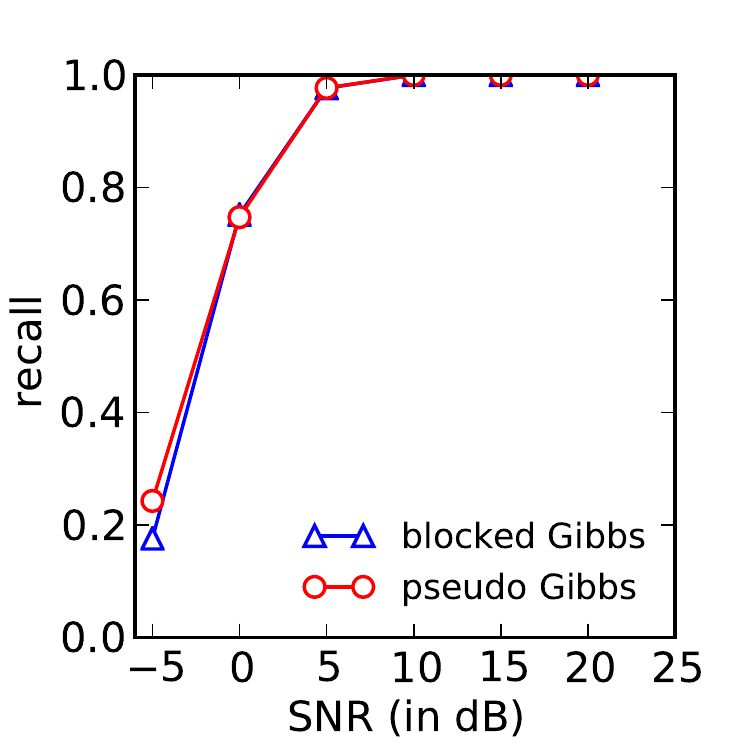}}
  \centerline{(b) Recall, $t=5$}\medskip
\end{minipage}
\begin{minipage}[b]{0.32\linewidth}
  \centering
  \centerline{\includegraphics[width=1.0\linewidth]{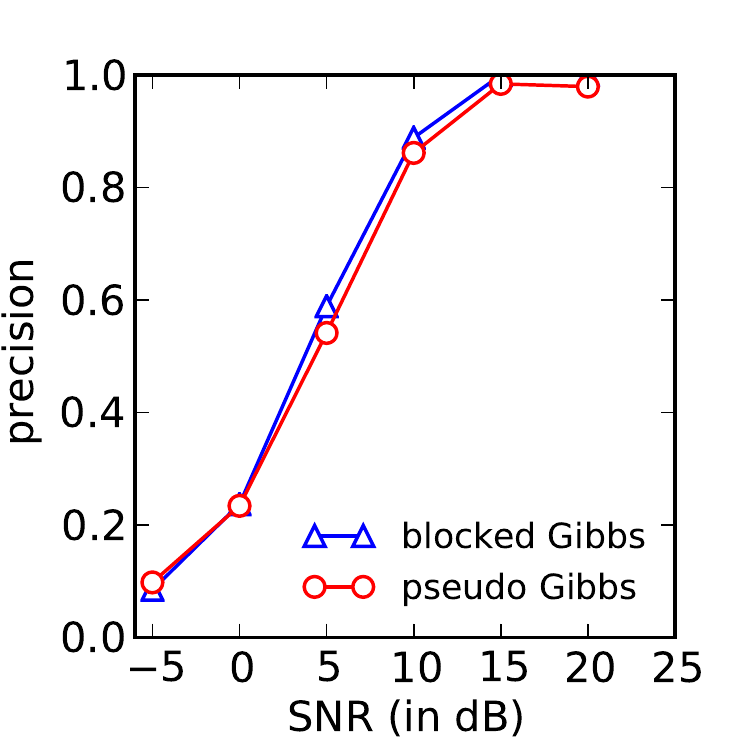}}
  \centerline{(d) Precision,  $t=0$}\medskip
\end{minipage}
\begin{minipage}[b]{0.32\linewidth}
  \centering
  \centerline{\includegraphics[width=1.0\linewidth]{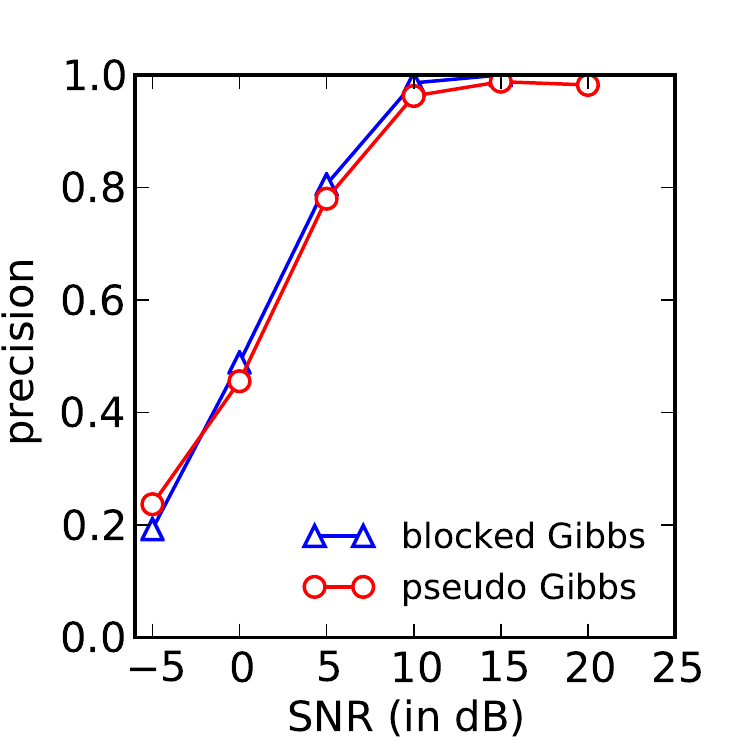}}
  \centerline{(e) Precision,  $t=1$}\medskip
\end{minipage}
\begin{minipage}[b]{0.32\linewidth}
  \centering
  \centerline{\includegraphics[width=1.0\linewidth]{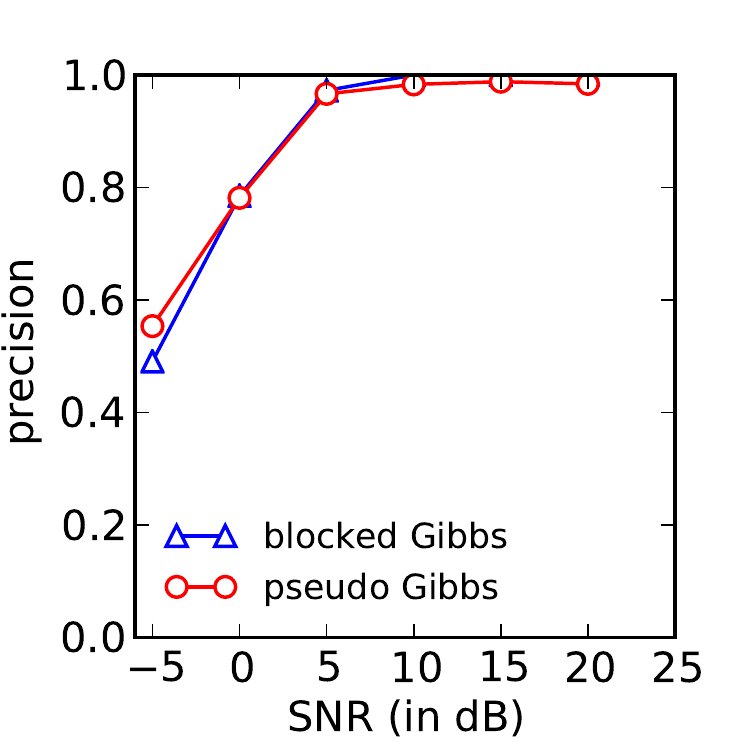}}
  \centerline{(f) Precision,  $t=5$}\medskip
\end{minipage}
\caption{Single change-point on data following a normal distribution with the Bernoulli detector model. Two cases are tested: with the blocked Gibbs sampler (blocked Gibbs, algorithm \ref{algo_uni_block}), and with the pseudo Gibbs sampler (pseudo Gibbs, algorithm \ref{algo_uni_pseudo}). The plots in (a), (b), (c) are recall curves and in (d), (e), (f) are precision curves, for several values of SNR. Three tolerances are tested in the estimated change-point location in time, at $\pm t$ points: (a), (d) $t=0$ (exact position), (b), (e) $t=1$ and (c), (f) $t=5$.}
\label{fig:rec_prec_Gibbs}
\end{figure}

Both algorithms with the blocked Gibbs sampler (algorithm \ref{algo_uni_block}) and the pseudo Gibbs sampler (algorithm \ref{algo_uni_pseudo}) are compared, for several values of the SNR. The acceptance level $\alpha$ is set to 0.01. 1000 MCMC iterations are done for each test. To quantify the performances of both algorithms in term of quality of the change-points detection, the precision and the recall are computed, for 1000 estimated MAP. These classical quantities are defined as:
\begin{equation}
recall = \frac{TP}{TP+FN}
\qquad
precision = \frac{TP}{TP+FP}
\end{equation}
where $TP$ is the number of true positive, $FN$ is the number of false negative and $FP$ is the number of false positive. The results for several SNR are shown in figure~\ref{fig:rec_prec_Gibbs}: the blocked Gibbs sampler and the pseudo-Gibbs sampler lead to the same detection performances, this justifies empirically the approximation done in algorithm~\ref{algo_uni_pseudo}. Due to the reduced computational cost of algorithm~\ref{algo_uni_pseudo}, we will use this algorithm, called Bernoulli detector, in the sequel of the paper.

\begin{figure}[htb]
\centering
\begin{minipage}[b]{0.32\linewidth}
  \centering
  \centerline{\includegraphics[width=1.0\linewidth]{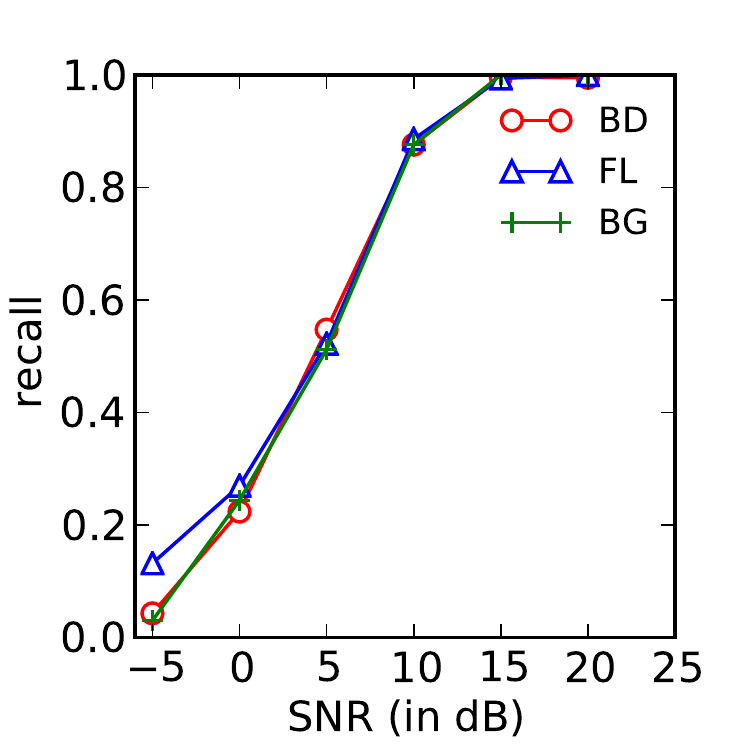}}
  \centerline{(a) Recall, $t=0$}\medskip
\end{minipage}
\begin{minipage}[b]{0.32\linewidth}
  \centering
  \centerline{\includegraphics[width=1.0\linewidth]{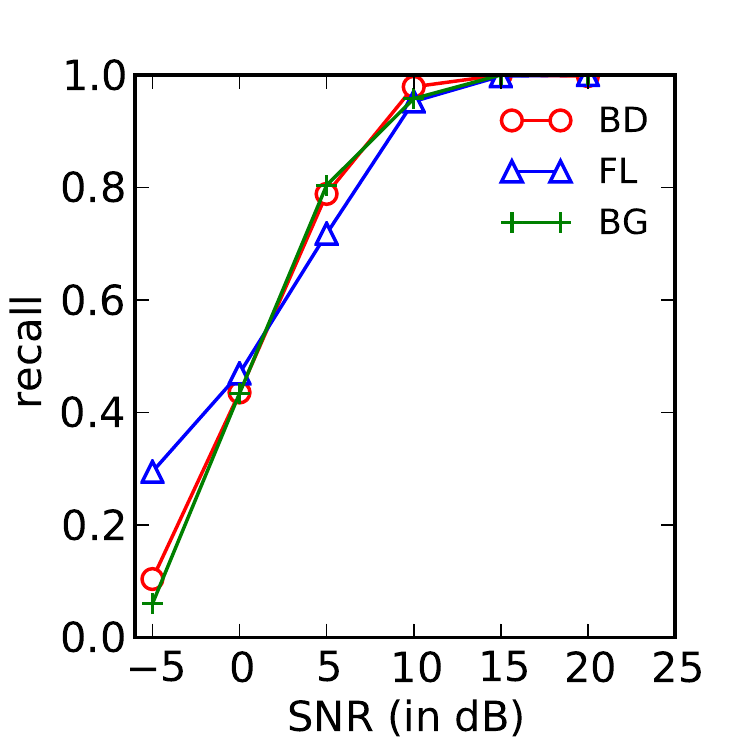}}
  \centerline{(b) Recall, $t=1$}\medskip
\end{minipage}
\begin{minipage}[b]{0.32\linewidth}
  \centering
  \centerline{\includegraphics[width=1.0\linewidth]{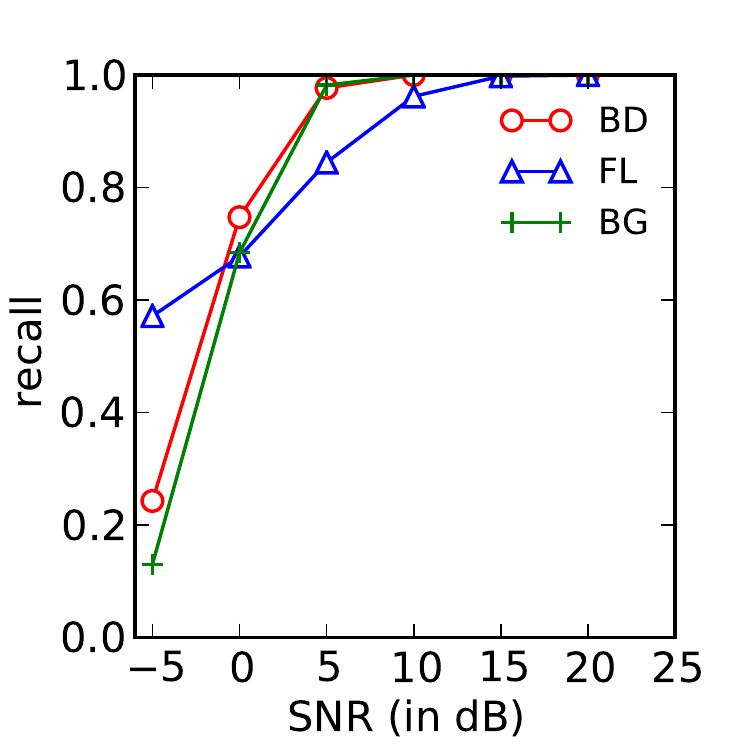}}
  \centerline{(b) Recall, $t=5$}\medskip
\end{minipage}
\begin{minipage}[b]{0.32\linewidth}
  \centering
  \centerline{\includegraphics[width=1.0\linewidth]{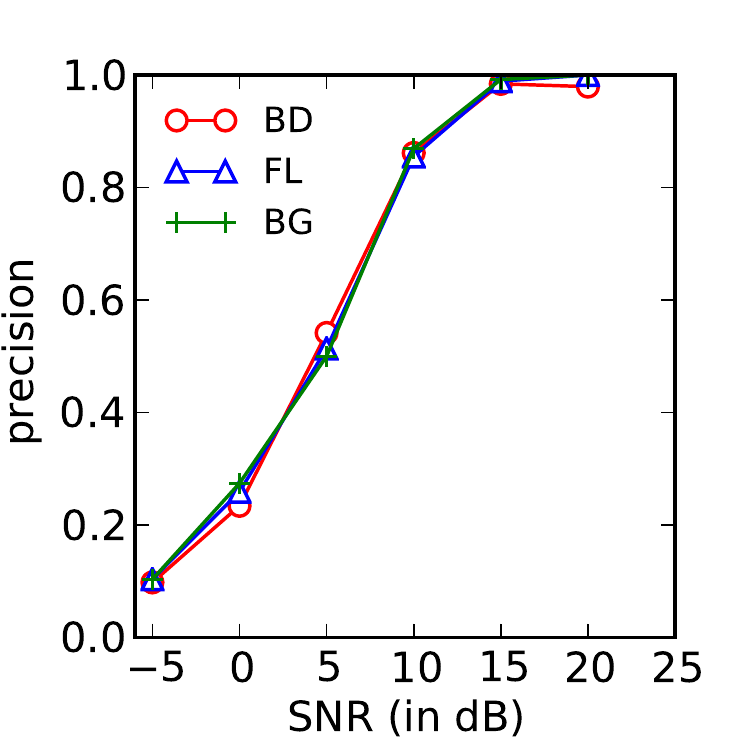}}
  \centerline{(d) Precision,  $t=0$}\medskip
\end{minipage}
\begin{minipage}[b]{0.32\linewidth}
  \centering
  \centerline{\includegraphics[width=1.0\linewidth]{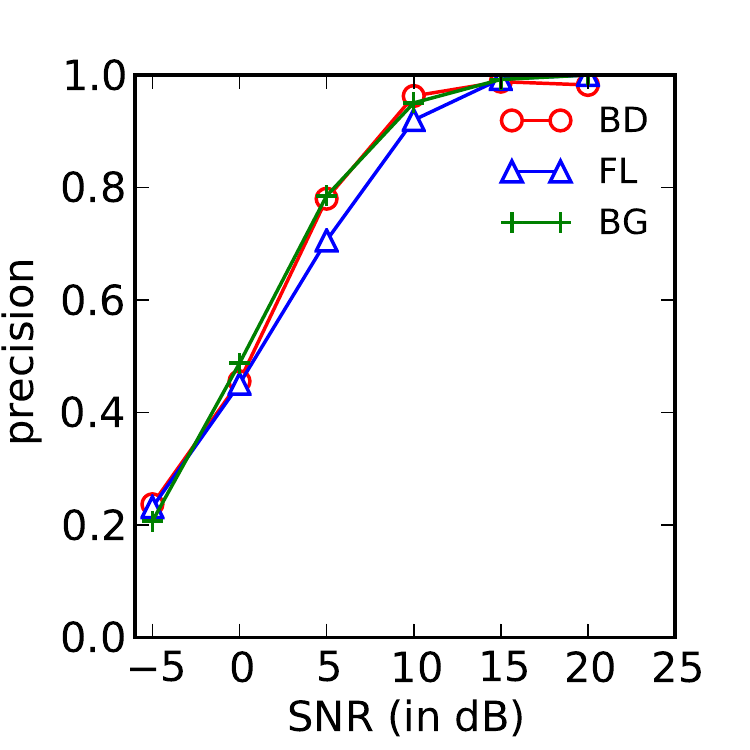}}
  \centerline{(e) Precision,  $t=1$}\medskip
\end{minipage}
\begin{minipage}[b]{0.32\linewidth}
  \centering
  \centerline{\includegraphics[width=1.0\linewidth]{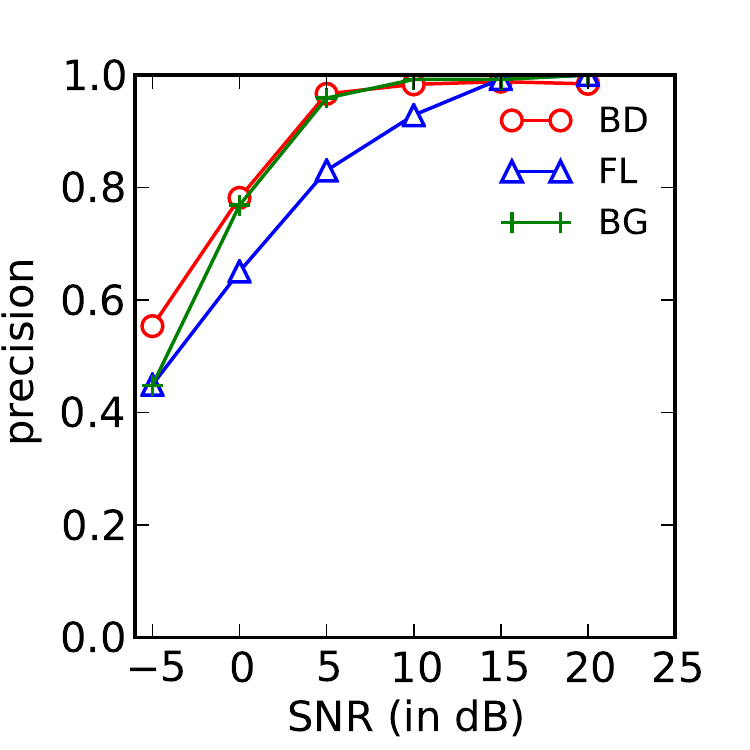}}
  \centerline{(f) Precision,  $t=5$}\medskip
\end{minipage}
\caption{Detection performances on data following a normal distribution of the Bernoulli detector model (BD), the fused lasso (FL) with parameter $\lambda=22.3$ and the Bernoulli Gaussian model (BG). The plots in (a), (b), (c) are recall curves and in (d), (e), (f) are precision curves, for several values of SNR. Several tolerances are tested in the estimated change-point location in time, at $\pm t$ points: (a), (d) $t=0$ (exact position), (b), (e) $t=1$ and (c), (f) $t=5$.}
\label{fig:rec-prec gaussian}
\end{figure}

The Bernoulli detector is compared to other classical approaches: the Bernoulli Gaussian model, and the fused lasso \citep{Tibshirani2011}. For this last one, we used the R package genlasso \citep{genlassoR}. The regularization parameter $\lambda$ has been adjusted to make the recall curve as similar as possible to the recall curve of the Bernoulli detector, for the exact change-point position. We choose $\lambda = 22.3$. 1000 tests are run for the Bernoulli detector (with 1000 MCMC iterations) and the fused lasso, and 500 for the Bernoulli Gaussian model (with 1000 MCMC iterations). The segmentation obtained with the fused lasso is determined from the coefficients of the solution of the optimization problem: a change-point is localized where the difference between two successive coefficients exceeds $10^{-10}$. The recall and precision curves for the three methods are plotted in figure~\ref{fig:rec-prec gaussian}. The recalls and the precisions are globally the same, except for the fused lasso when the 
tolerance 
in the change-point position is $\pm 5$ time points. This is explained by the fact that the fused lasso estimates a few more change-points than the other methods for the $\lambda$ we choose. However this method is parametrizable by $\lambda$ and is much faster than the other ones with the MCMC approach.\\

\setcounter{subsection}{1}
\refstepcounter{subsection}
\noindent {\bf 2.6.2. False discovery rate}
\label{ssec:FDR}

To maximize the probability of detecting the true positive by controlling the false positives in the case of multiple hypothesis testing, 
a widely used way is to control the false discovery rate (FDR). 
This term is defined as the expected proportion of incorrectly rejected null hypotheses:
\begin{equation}
FDR = E \Big[ \frac{V}{R\vee1}\Big]
\label{FDR}
\end{equation}
where $V$ is the number of false positives, and $R$ is the number of positives.
When the $m$ tests are independent, the Benjamini-Hochberg procedure, presented in~\citep{Benjamini1995}, allows to control the FDR. 
However the case of dependent tests statistics are frequently encountered, that is why other procedures have been proposed to control this rate. 
In the case where the test statistics are positively dependent, the authors of~\citep{Benjamini2001} present another procedure. 
In our case, the $p$-values computed by the statistical test are highly dependent, because the positions of the change-points define the 
segments for the test. The addition or the suppression of one of them affects the previous and the next $p$-value. 
Despite this particular dependency, we are interested in the evolution of the FDR with the acceptance level $\alpha$.

\begin{figure}[htb]
\begin{center}
\includegraphics[width=0.6\linewidth]{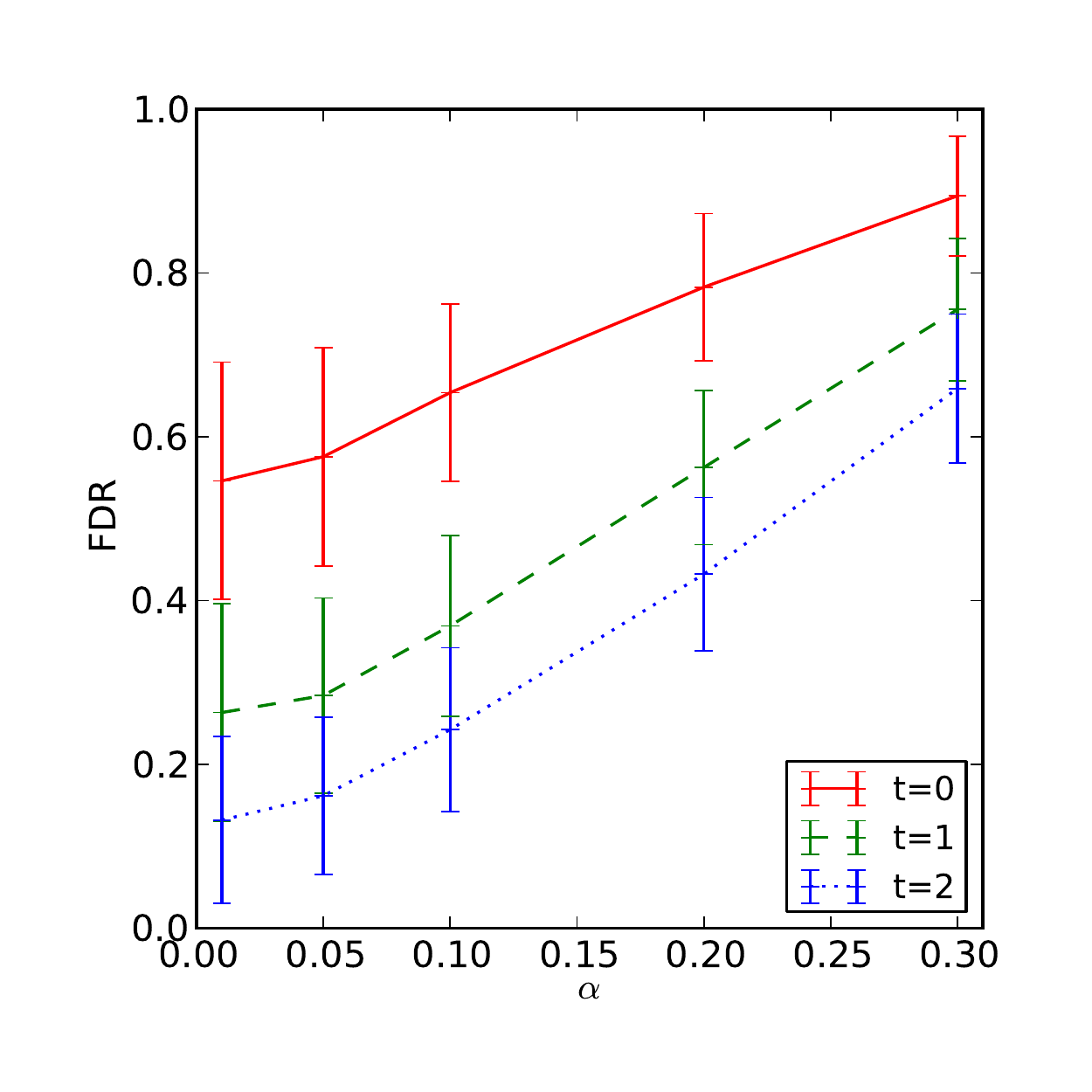}
\caption{Evolution of the FDR with $\alpha$ and its standard deviation, for several precisions $t$ in the position of the estimated change-points.}
\label{fig:FDR}
\end{center}
\end{figure}

Tests have been conducted on a signal of $N=320$ time points, made of 16 segments of 20 points. 
The data are generated from a normal distribution. Between two successive segments, the difference 
between the means is $\pm 1.0$ and the SNR, 
as defined for the previous simulation~\ref{ssec:single_CP_uni}, is 5.0 dB. The FDR is computed from 
the mean of the ratios $\frac{V}{R}$ (or 0 if $R=0$) obtained from the MAP estimators of 350 simulations with 2000 MCMC iterations, 
for different values of $\alpha$. There are $m=318$ hypothesis to test, and 15 of them are real discoveries. 
Note that the first and the last points of the signal are not included. The results are shown in figure~\ref{fig:FDR},
for three tolerances in the estimated change-points positions. 
The FDR is increasing with the acceptance level $\alpha$. This empirically confirms a global control 
by $\alpha$ on the change-points detection, 
in addition to the single change-point control detailed in proposition \ref{prop:MAP} and \ref{prop:MMSE}.
\\

\setcounter{subsection}{2}
\refstepcounter{subsection}
\noindent {\bf 2.6.3. Robustness with respect to outliers}
\label{ssec:outliers_uni}

\begin{figure}[htb]
\centering
\begin{minipage}[b]{0.32\linewidth}
  \centering
  \centerline{\includegraphics[width=1.0\linewidth]{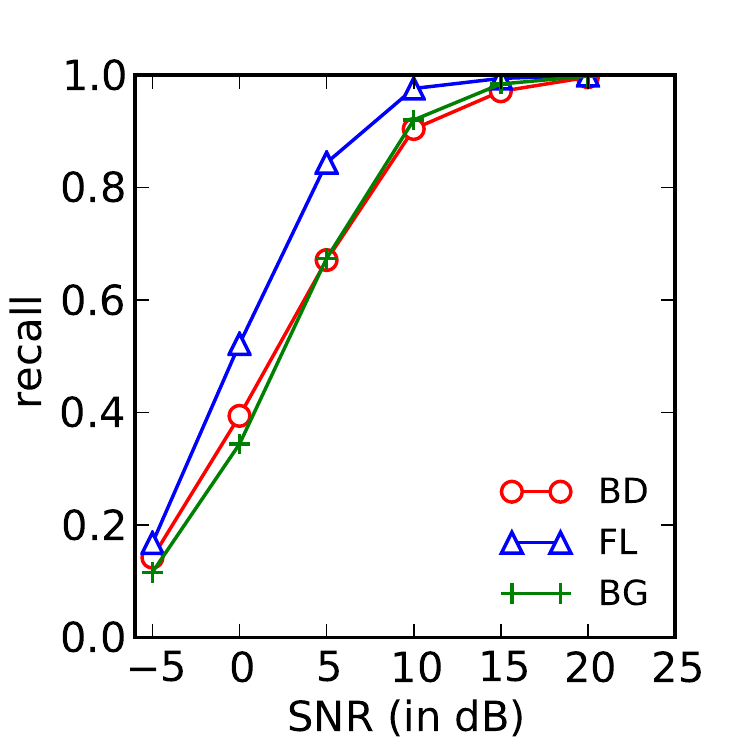}}
  \centerline{(a) Recall, $t=0$}\medskip
\end{minipage}
\begin{minipage}[b]{0.32\linewidth}
  \centering
  \centerline{\includegraphics[width=1.0\linewidth]{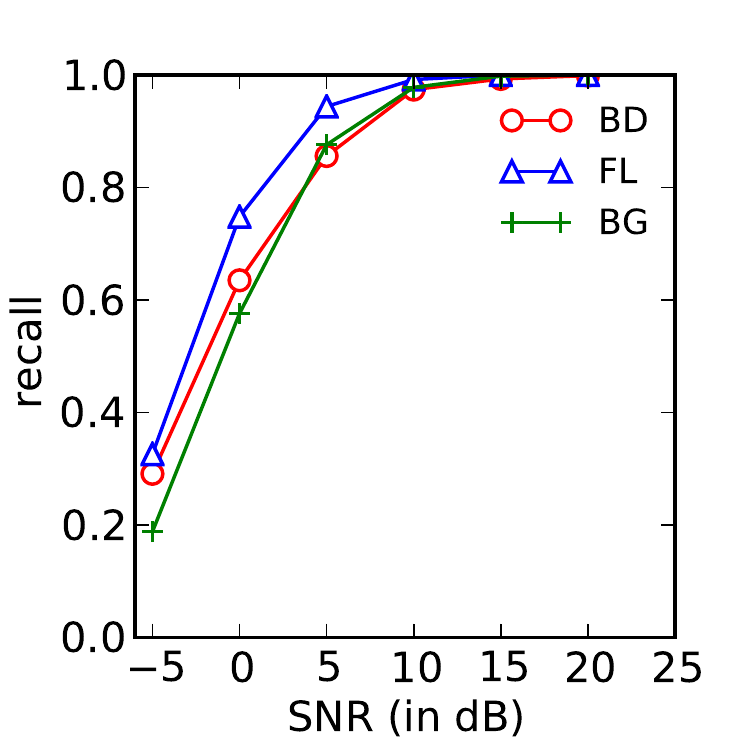}}
  \centerline{(b) Recall, $t=1$}\medskip
\end{minipage}
\begin{minipage}[b]{0.32\linewidth}
  \centering
  \centerline{\includegraphics[width=1.0\linewidth]{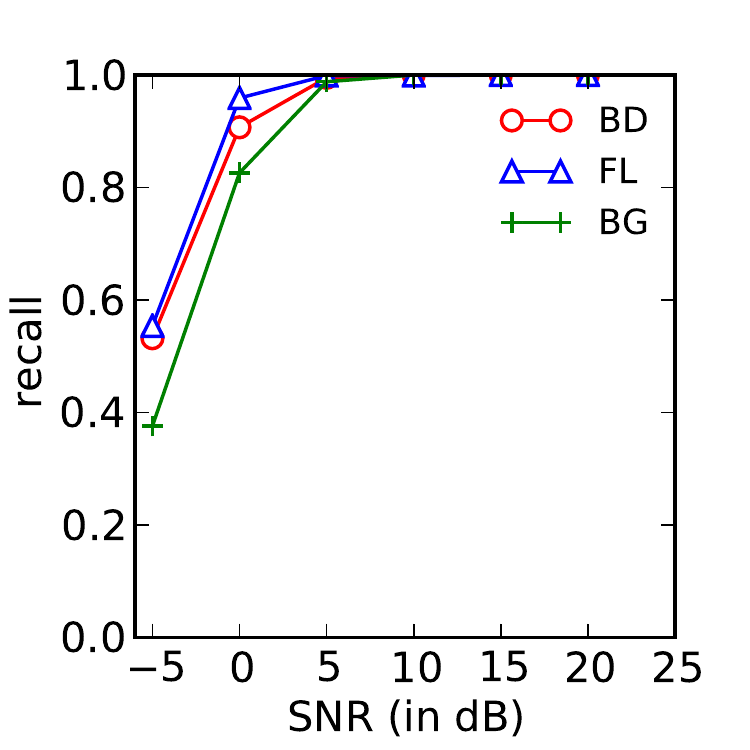}}
  \centerline{(b) Recall, $t=5$}\medskip
\end{minipage}
\begin{minipage}[b]{0.32\linewidth}
  \centering
  \centerline{\includegraphics[width=1.0\linewidth]{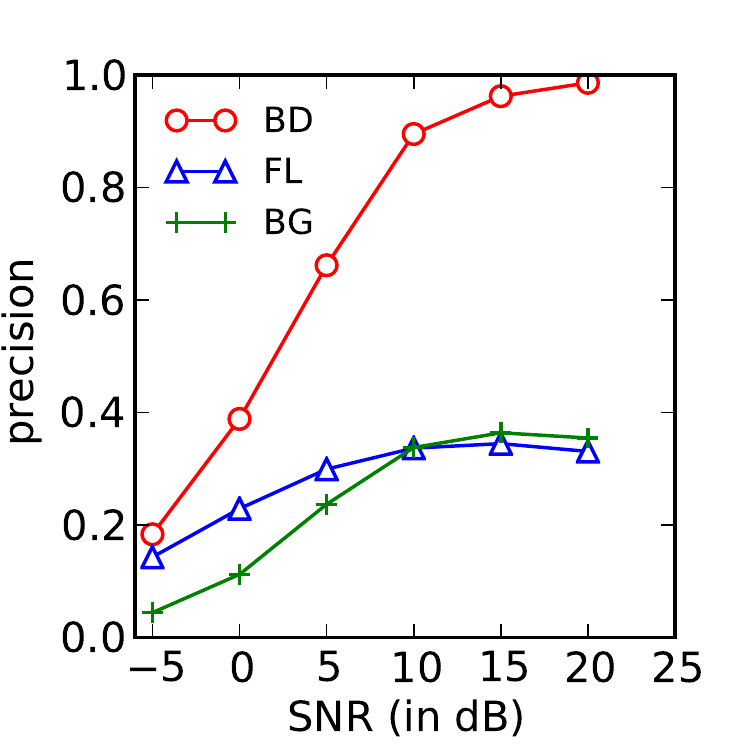}}
  \centerline{(d) Precision,  $t=0$}\medskip
\end{minipage}
\begin{minipage}[b]{0.32\linewidth}
  \centering
  \centerline{\includegraphics[width=1.0\linewidth]{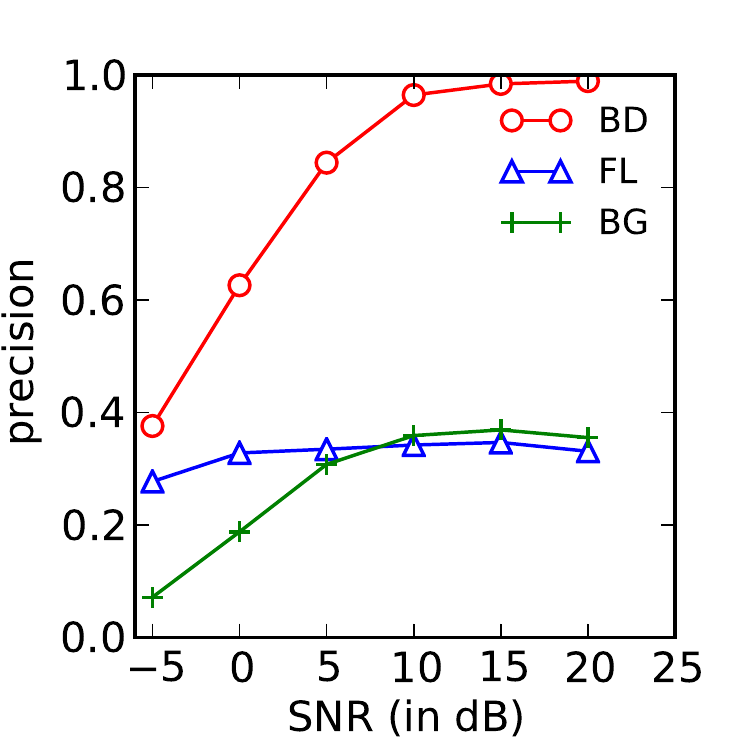}}
  \centerline{(e) Precision,  $t=1$}\medskip
\end{minipage}
\begin{minipage}[b]{0.32\linewidth}
  \centering
  \centerline{\includegraphics[width=1.0\linewidth]{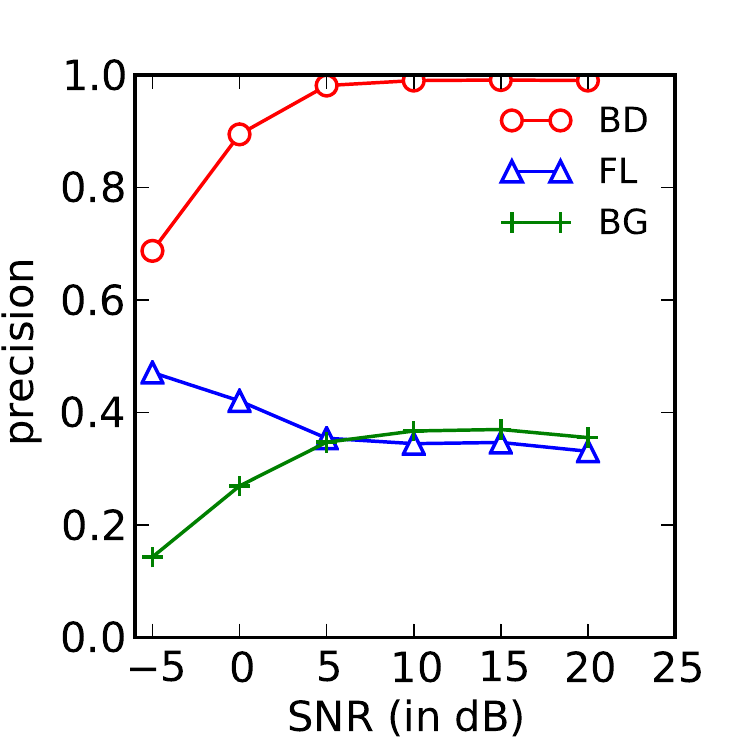}}
  \centerline{(f) Precision,  $t=5$}\medskip
\end{minipage}
\caption{Detection performances on data following a Student's t-distribution of the Bernoulli detector model (BD), the fused lasso (FL) with parameter $\lambda=22.3$ and the Bernoulli Gaussian model (BG). The plots in (a), (b), (c) are recall curves and in (d), (e), (f) are precision curves, for several values of SNR. Several tolerances are tested in the estimated change-point location in time, at $\pm t$ points: (a), (d) $t=0$ (exact position), (b), (e) $t=1$ and (c), (f) $t=5$.}
\label{fig:rec-prec student}
\end{figure}

In this experiment, the ability to detect change-points in data with outliers is tested. Several signals are generated with a single change-point, similarly as in section~\ref{ssec:single_CP_uni}, but here the observations of the segment $k$ follow a Student's t-distribution of parameters $(\nu,\mu_k,\sigma)$, where $\nu=3.0$ and $\sigma^2 = \frac{\nu}{\nu-2}$. The SNR is defined between two segments, by the expression~\eqref{eq:SNR}. The Bernoulli detector model is applied with an acceptance level $\alpha=0.01$, and 1000 MCMC iterations are done. The recall and precision are computed from 1000 MAP estimators of $R$, for several values of SNR. The curves are presented in the figure~\ref{fig:rec-prec student}, for three tolerances in the estimated change-point position.

To compare these results, fused lasso (1000 simulations) and the classical Bernoulli Gaussian model (500 simulations) are applied on the same data. the regularization parameter $\lambda$ is set to the same value as for the normal data. The resulting recalls and precisions 
are depicted in the figure~\ref{fig:rec-prec student}. 
Like in the normal case, the recall curves are similar, but there are significant differences between the methods 
for the precision. The Bernoulli detector's performances do not change with the data following a Student's t-distribution.
But the fused lasso and the Bernoulli Gaussian model detect several false change-points, due to the presence of outliers. 
This result shows that when the Gaussian observation model is very different from the distribution of the data, 
it should be changed to be closer to the correct distribution. 
For this particular simulation, the observation model should have been taken for distributions with heavy tails. 
Our approach presents the advantage to be adequate to a full range of data distributions having good performances either for Gaussian distributions or distributions with heavy tail.   
\\

\setcounter{chapter}{2}
\refstepcounter{chapter}
\setcounter{equation}{2}
\noindent {\bf 3. Extension to multivariate case}
\label{chap:multi}

The Bernoulli detector model can be extended to the multivariate case. Then, the data of the $K$ time series are 
stored in a $K\times N$ matrix $\bX$, where each signal is a row of $\bX$, and the indicators matrix $\bR$ is 
a $K\times N$ matrix whose rows are the one defined in equation~(\ref{def R}), with the convention 
that $r_{j,1}=r_{j,N}=1$, $\forall j \in \{1,...,K\}$. Like in the univariate case, we are interested in 
the comparison of the medians of two successive segments $s^+$ and $s^-$, to apply the rank test. 
The observations are simply supposed to be independent in time and from a signal to another.

In \citep{YutFong2011}, the authors introduce a method for multiple change-points detection in multivariate time series, 
based on a multivariate extension of the Wilcoxon rank sum test.  However in that case, the change-points are supposed 
to be simultaneous across all time series. It leads to a more powerful test than some independent univariate approaches 
as soon as the joint change-points assumption is valid. In the case of real data, this assumption often appears to be too 
restrictive. 

In our model assumption, the change-points are not necessarily shared
across the whole set of time series. The column vector $R_i = (r_{1,i},...,r_{K,i})^T$, for signals $1$ to $K$, gives the configuration 
$R_i \in \{0,1\}^{K}$ at time index $i$. 
This configuration, written $\epsilon$, denotes which time series have a simultaneous change-point at a given time $i$. 
$\mathcal{E}$ is the set of all possible configurations, where $\mathcal E$ is a subset of $\{0, 1\}^K$. Without any 
assumption on the possible configurations, that is in a noninformative case, $\mathcal E = \{0, 1\}^K$. A parameter $\bP$, 
defined as the vector of probabilities $P_\epsilon$, is introduced to denote the probabilities to have each of all possible 
configurations $\epsilon$ in the indicators matrix $\bR$. The Bayes' rule now yields:
\begin{equation}
f(\bR|\bX) \propto \int_\bP L_{\ast}(\bX|\bR) f(\bR|\bP) f(\bP) d\bP.
\label{bayes P}
\end{equation}
The specific prior chosen for the parameter $\bP$ is detailed in part~\ref{sec:prior_multi}.
\\

\setcounter{section}{0}
\refstepcounter{section}
\noindent {\bf 3.1. Change-point model}
\label{sec:CP_model_multi}

The likelihood function is built like in the univariate case for each signal.
As the observations of the matrix $\bX$ are assumed to be independent, the likelihood expresses as follow:
\begin{equation}
L(\bX|\bR) = \prod_{j=1}^{K} \prod_{i=2}^{N-1} f (x_{j,i} | r_{j,i}).
\end{equation}
Then, the $p$-value $p_{j,i}$ for $x_{j,i}$ is computed by considering the previous and next segment in the signal $j$: 
$s_{j,i^-} = (x_{j,k})_{i^-+1 \leq k \leq i}$ and $s_{j,i^+} = (x_{j,k})_{i+1 \leq k \leq i^+}$, where $i^-$ and $i^+$ 
are the indexes of the previous and the next change-points respectively around $x_{j,i}$ in the signal $j$, 
given the indicators matrix $\bR$. The same parametrization of the $p$-values as in (\ref{eq:BetaPar}) and (\ref{eq:densP}) is chosen. 
It leads finally to the inference function
\begin{align}
L_{\ast}(\bX|\bR)& = \prod_{j=1}^{K} \prod_{i=2}^{N-1} \left( \gamma p_{j,i}^{\gamma-1} \right)^{r_{j,i}}.
\label{eq:L_multi}
\end{align}
\\

\setcounter{section}{1}
\refstepcounter{section}
\noindent {\bf 3.2. Prior on indicators}
\label{sec:prior_multi}

The time series have a dependency structure that depicts the probabilities that the same event impacts the connected signals at 
the same time. Here the events are localized as change-points. The connected signals, for example those that are linked by a 
dependency relationship, have then a high probability to present their change-points simultaneously. With our notations, 
it means that if the signal $k$ depends on the signal $l$, then $R_{k,i} = R_{l,i}$ with a high probability, and the coefficients 
$\epsilon(k)$ and $\epsilon(l)$ tend to have the same value for a given time point. Following the approach presented in~\citep{DobigeonApril}, 
this leads to consider the vector of probabilities $\bP = (P_\epsilon)_{\epsilon \in \mathcal{E}}$, 
where $P_\epsilon$ is the probability to observe the configuration $\epsilon$ in $\bR$. 
The parameter $\bP$ encodes the dependencies of the change-point between the different time series.
This gives us an information about the dependency structure between the time series.

In the multivariate case, the vectors $(R_i)_{2 \le i \le N-1}$ are assumed to be \textit{a priori} independent, 
thus the equation~(\ref{eq:f(R|q)}) for the prior on indicators becomes
\begin{equation}
f(\bR) = \prod_{i=2}^{N-1} f(R_i).
\label{eq:prior_multi}
\end{equation}

When the vector $\bP$ of probabilities of the configurations is explicitly expressed, the equation 
(\ref{eq:prior_multi}) becomes
\begin{equation}
f(\bR|\bP) = \prod_{\epsilon \in \mathcal{E}} P_\epsilon^{S_\epsilon(\bR)}
\label{eq:priorPe}
\end{equation}
where $S_\epsilon(\bR)$ is the number of times that the configuration $\epsilon$ appears in the columns of $\bR$.

As in~\citep{DobigeonApril}, a vague prior is chosen for $\bP$: the parameter follows a Dirichlet distribution 
$\mathcal{D}_{L}(d)$, with hyperparameter vector $d= (d_\epsilon)_{\epsilon \in \mathcal{E}}$ and where $L$ is the 
cardinal of $\mathcal E$. All the $d_\epsilon$ are set to the same deterministic value $d_\epsilon \equiv d=1$, 
then the distribution of $\bP$ is uniform.
\\

\setcounter{section}{2}
\refstepcounter{section}
\noindent {\bf 3.3. Posterior distribution}
\label{sec:posterior_multi}

The relations (\ref{eq:L_multi}), (\ref{eq:priorPe}) and the choice of the prior on the hyperparameter $\bP$ leads to the following posterior:
\begin{align}
f(\bR,\bP|\bX)& \propto  L_{\ast}(\bX|\bR)  f(\bR|\bP) f(\bP), \\
&\propto \left(  \prod_{j=1}^{K} \prod_{i=2}^{N-1} \left( \gamma p_{j,i}^{\gamma-1} \right)^{r_{j,i}} \right) 
\left( \prod_{\epsilon \in \mathcal{E}} P_\epsilon^{S_\epsilon(\bR) + d_\epsilon -1} \right).
\label{eq:postR,P|X}
\end{align}
The vector of parameters $P_\epsilon$ can be integrated out. Since $\bP$ follows the Dirichlet distribution
\begin{equation}
P_\epsilon \sim \mathcal{D}_L(S_\epsilon(\bR) + d_\epsilon),
\label{eq:P|Se,d}
\end{equation}
the marginalized posterior becomes:
\begin{equation}
f(\bR|\bX) \propto \left( \prod_{j=1}^{K} \prod_{i=2}^{N-1} (\gamma\pji^{\gamma-1})^{\rji} \right) 
\times \frac{\prod_{\epsilon \in \mathcal E} \Gamma (S_\epsilon(\bR)+d_\epsilon)} {\Gamma (N+L)}.
\label{eq:post_multi}
\end{equation}
\\

\setcounter{section}{3}
\refstepcounter{section}
\setcounter{algocf}{2}
\noindent {\bf 3.4. Algorithm}
\label{sec:algo_multi}

The pseudo-code of the new method is described in the algorithm~\ref{algo_multi}. Again, a MCMC method is applied with a Gibbs sampling strategy using the pseudo Gibbs approximation. It allows here, when $r_{j,i}=1$, to avoid the sampling of each combinations of 0 and 1 for a block of indicators at time indexes $i-1$, $i$, $i+1$ and for all $J$ signals, that is $(2^3)^K$ configurations. The columns vectors $R_i$ are sampled at each iteration, following the posterior~\eqref{eq:post_multi}, by updating only the $p$-values $(p_{j,i})_{1 \leq j \leq K, i}$. An additional step can be done to sample the vector $\bP$ in order to have the distribution of each configuration $\epsilon \in \mathcal{E}$.
The complexity of this multivariate algorithm depends linearly on the number $L$ of tested configurations $\epsilon$ 
for the columns of $\bR$.
\begin{algorithm}
\caption{Multivariate Bernoulli-Detector}
\label{algo_multi}
\SetKw{Req}{\textbf{require}} \Req{$\mathcal E = \{\epsilon_0,...,\epsilon_l,...,\epsilon_L\} \subset \{0,1\}^J$, $\alpha$} \\
\SetKw{Ini}{\textbf{initialize}} \Ini{$\bR^{(0)}$, $S_\epsilon(\bR^{(0)})$, $M$}\\
\For{$m\leftarrow 1$ \KwTo $M$}{
	initialize the index set $I = \{2,\ldots,N-1\}$ \\
	\While{$I \neq \emptyset $}{
		pick randomly $i$ in $I$ \\
		\For{$j\leftarrow 1$ \KwTo $K$}{
			compute $p_{j,i}^{(m)}$
		}
		sample $R_i^{(m)}$ from its posterior \eqref{eq:post_multi}\\
		remove $i$ from $I$\\
	}
	\textit{optional} sample $\bP^{(m)}$ from its posterior \eqref{eq:P|Se,d}\\
}
\Return{$\bR$, $\bP$}\\
\end{algorithm}
\\

\setcounter{section}{4}
\refstepcounter{section}
\noindent {\bf 3.5. Simulations}
\label{sec:simu_multi}

\begin{figure}[htb]
\begin{center}
\includegraphics[width=0.25\linewidth,trim = 6mm 16mm 10mm 8mm,clip]{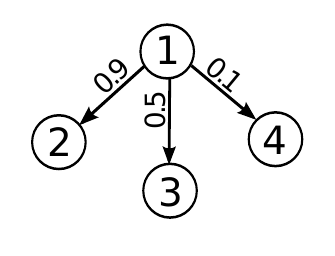}
\vspace*{-0.5cm}
\caption{Dependency structure between the four simulated time series for the existence of change-point. Each node represents a signal. The weight $w_{k \rightarrow l}$ of the directed edge from node $k$ to $l$ denotes the probability that a change-point in signal $k$ exists in signal $l$ simultaneously.}
\label{fig:simu_multi_dep_struct}
\end{center}
\end{figure}

\begin{figure}[htb]
\begin{center}
\includegraphics[width=1.0\linewidth]{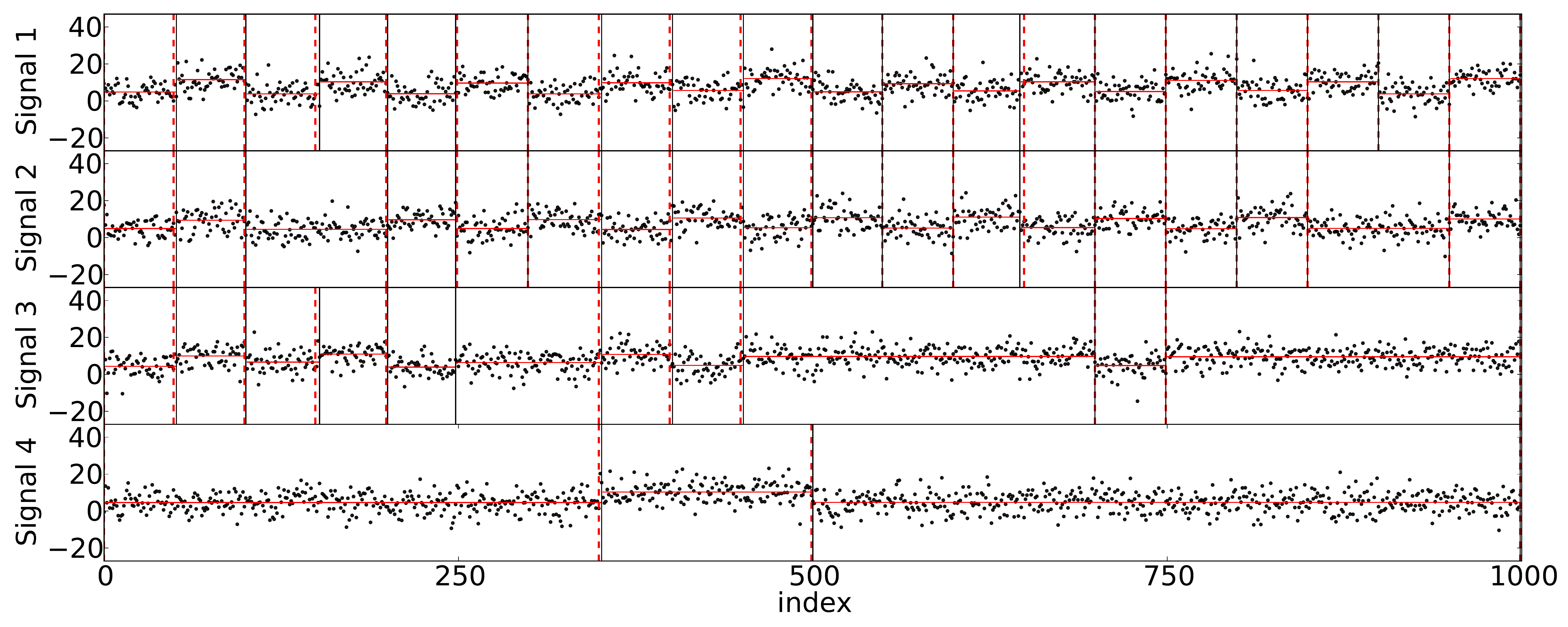}
\caption{Joint detection of change-points in multivariate time series. The true and estimated change-points positions are shown by the dashed and full vertical lines respectively. The means of each segment is added in full line.}
\label{fig:simu_multi_MAP}
\end{center}
\end{figure}

In this experiment, we simulated four time series of 1000 points, with the dependency structure shown in figure~\ref{fig:simu_multi_dep_struct}: all change-points are induced by signal 1, with different probabilities. There are 20 segments in signal 1. Within a segment $s$, all observations are i.i.d. and follow a normal distribution $\mathcal{N}(\mu_s,\sigma)$, such that between two successive segments $s^+,s^-$, the SNR, defined in equation~(\ref{eq:SNR}), is 0.0 dB. The algorithm is applied with an noninformative prior on the dependencies between time series: namely all configurations $\epsilon$ are tested, $\mathcal{E} = \{0,1\}^4$. The acceptance level $\alpha$ is 0.01, and 2000 MCMC iterations are done. 
The resulting MAP estimation of the change-points positions is represented in figure~\ref{fig:simu_multi_MAP}. 
This simulation is showing that the change-points are well estimated, and those that are not precisely at their true position are though simultaneously detected on the expected signals (see 
for 
instance
the change-point at time 60, on signals 1 and 3).

\begin{figure}[htb]
\begin{center}
\includegraphics[width=0.8\linewidth]{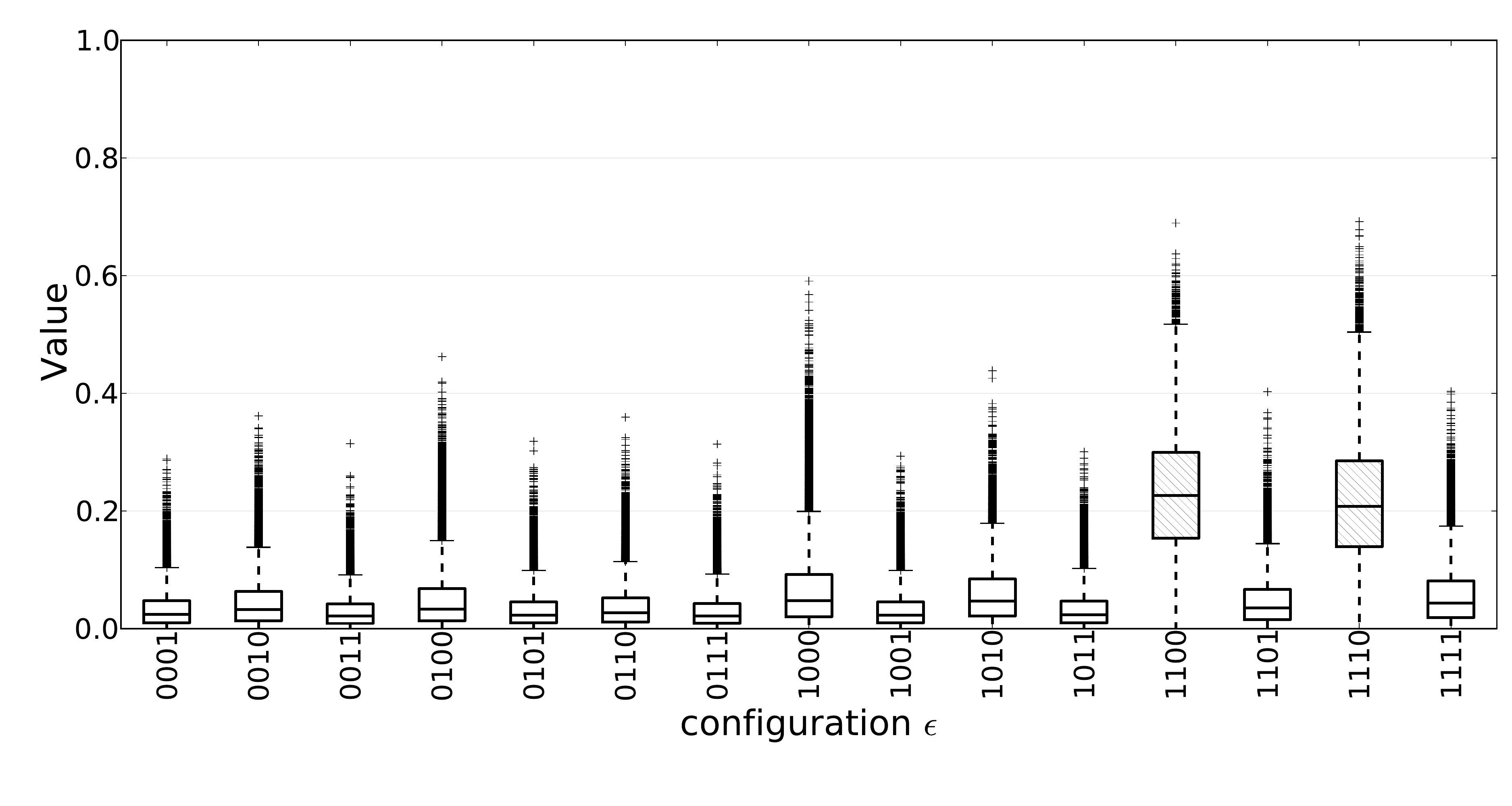}
\caption{Posterior distributions of the $P_\epsilon$, for each configuration $\epsilon$ in $\mathcal{E} = \{0,1\}^4$, in multivariate time series. The most important ones are hatched.}
\label{fig:simu_multi_Peps}
\end{center}
\end{figure}

The probabilities $P_\epsilon$ are drawn, following the distribution (\ref{eq:P|Se,d}), to show the estimation of the dependency structure. They are represented in figure~\ref{fig:simu_multi_Peps}. To increase the values, the conditional  distributions are computed, given the case that there is at least one change-point. In figure~\ref{fig:simu_multi_Peps}, there is one boxplot by distribution of probability $P_\epsilon$. The most important ones are $P_{1100}$ and $P_{1110}$, due to the higher probabilities to have the same change-points in signal 1 and 2 and in signal 1, 2 and 3 simultaneously.
\\

\setcounter{chapter}{3}
\refstepcounter{chapter}
\setcounter{equation}{3}
\noindent {\bf 4. Applications}
\label{chap:appli_multi}

\setcounter{section}{0}
\refstepcounter{section}
\noindent {\bf 4.1. Household electrical power consumption}
\label{sec:elec}

In this section, the ability of the algorithm to learn the dependency
structure from data is illustrated using measurements of household power
consumption~\citep{Bache+Lichman:2013}. This real dataset consists of four
time series.
One of them, denoted 1, depicts the global electrical energy consumption in
the house, while the others (respectively denoted 2, 3 and 4) are devoted to
the measurements of power demand by specific devices. Thus, the relationships
between the four time series is known: the signals 2, 3 and 4 are independent
of each other, and the signal 1 is the sum of the signals 2, 3 and 4, and of
other missing signals.

\begin{figure}[htb]
\begin{center}
\includegraphics[width=1.0\linewidth]{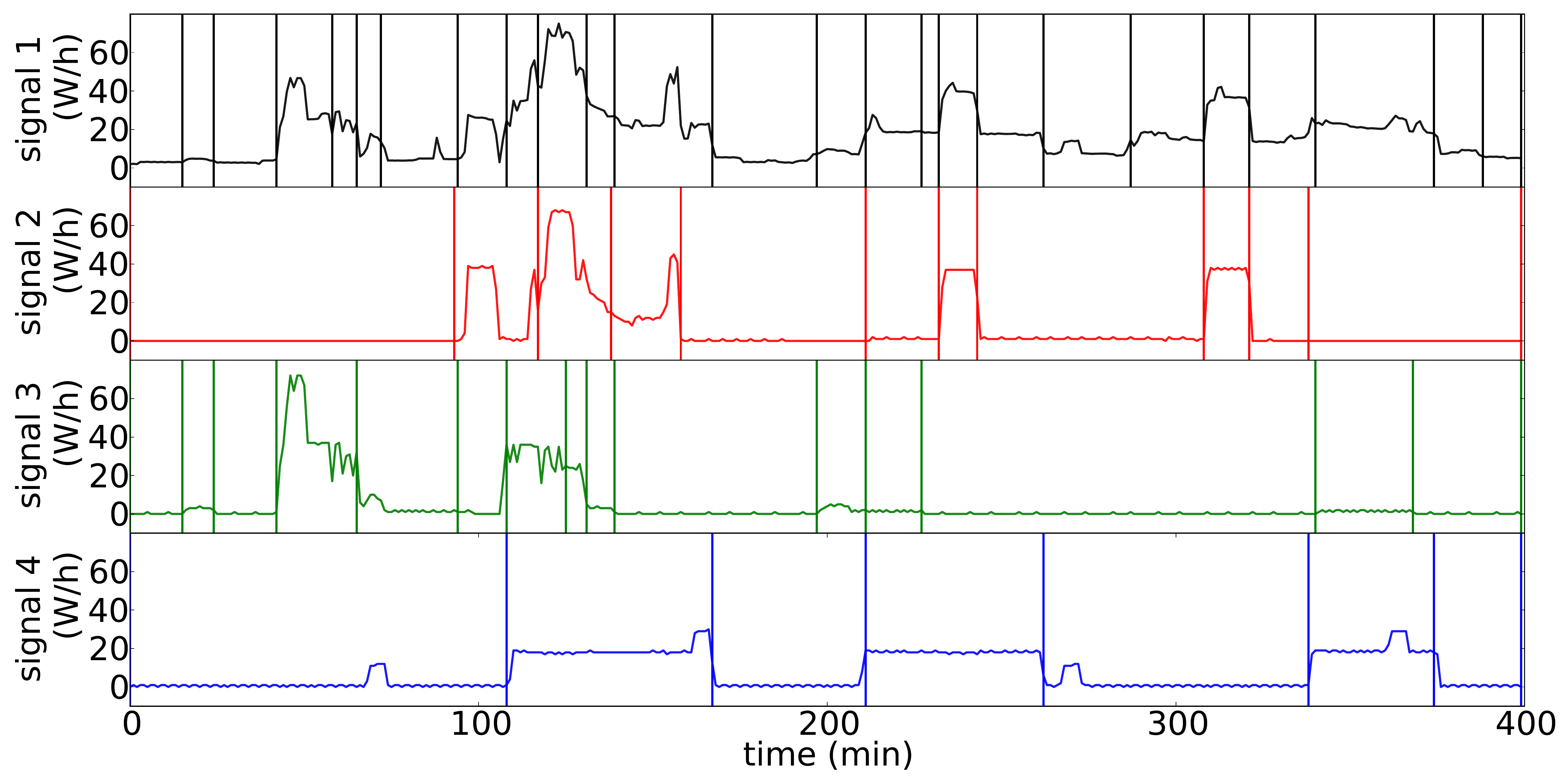}
\caption{Segmentation of the household electrical power consumption dataset, with a noninformative prior. The detected change-points are the vertical lines.}
\label{fig:MAP_elec}
\end{center}
\end{figure}

\begin{figure}[htb]
\begin{center}
\includegraphics[width=0.8\linewidth]{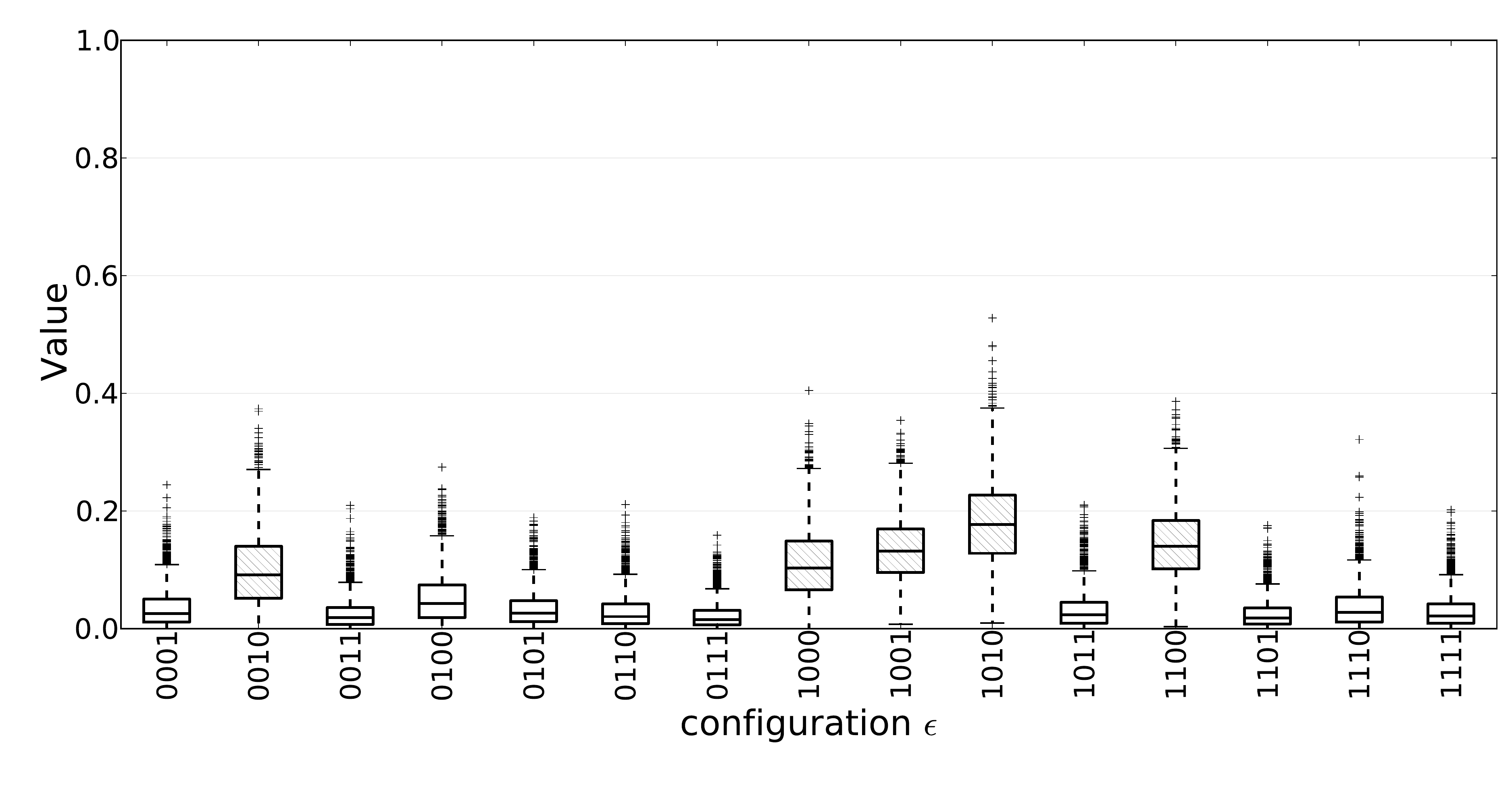}
\caption{Posterior distributions of the $P_\epsilon$, for the household electrical power consumption dataset, with a noninformative prior. The most important ones are hatched.}
\label{fig:Peps_elec}
\end{center}
\end{figure}

The Bernoulli detector model is applied with a noninformative prior on the
dependency structure, for 2000 MCMC iterations. Visual results on a small
portion of signals are represented in figure~\ref{fig:MAP_elec} for the
change-points detection: the major changes are detected. Similarly to the previous
simulations, the probabilities of each non-empty configuration $\epsilon$,
given that there is at least one change-point, are plotted in
figure~\ref{fig:Peps_elec}. The most probable configurations in $R$ are, in
decreasing order of the median, 1010, 1100, 1001, 1000 and 0010. The three
first ones reveal the links within the time series of the groups $(1,3)$,
$(1,2)$ and $(1,4)$. The importance of $P_{1000}$ is due to the unknown events
from the hidden part of the electrical installation, not measured. The signal
3 seems to have several change-points that are not found in signal 1. The
choice of an informative prior on the configurations would eliminate this
problem: only the configurations such that if $\epsilon(k) = 1$ then
$\epsilon(1) = 1$ for $k \in \{2,3,4\}$ are allowed. This would also help to
find a better time localization for some estimated change-points.
\\

\setcounter{section}{1}
\refstepcounter{section}
\noindent {\bf 4.2. Comparative Genomic Hybridization array data}
\label{sec:aCGH}

\begin{figure}[htb]
\begin{center}
\includegraphics[width=1.0\linewidth]{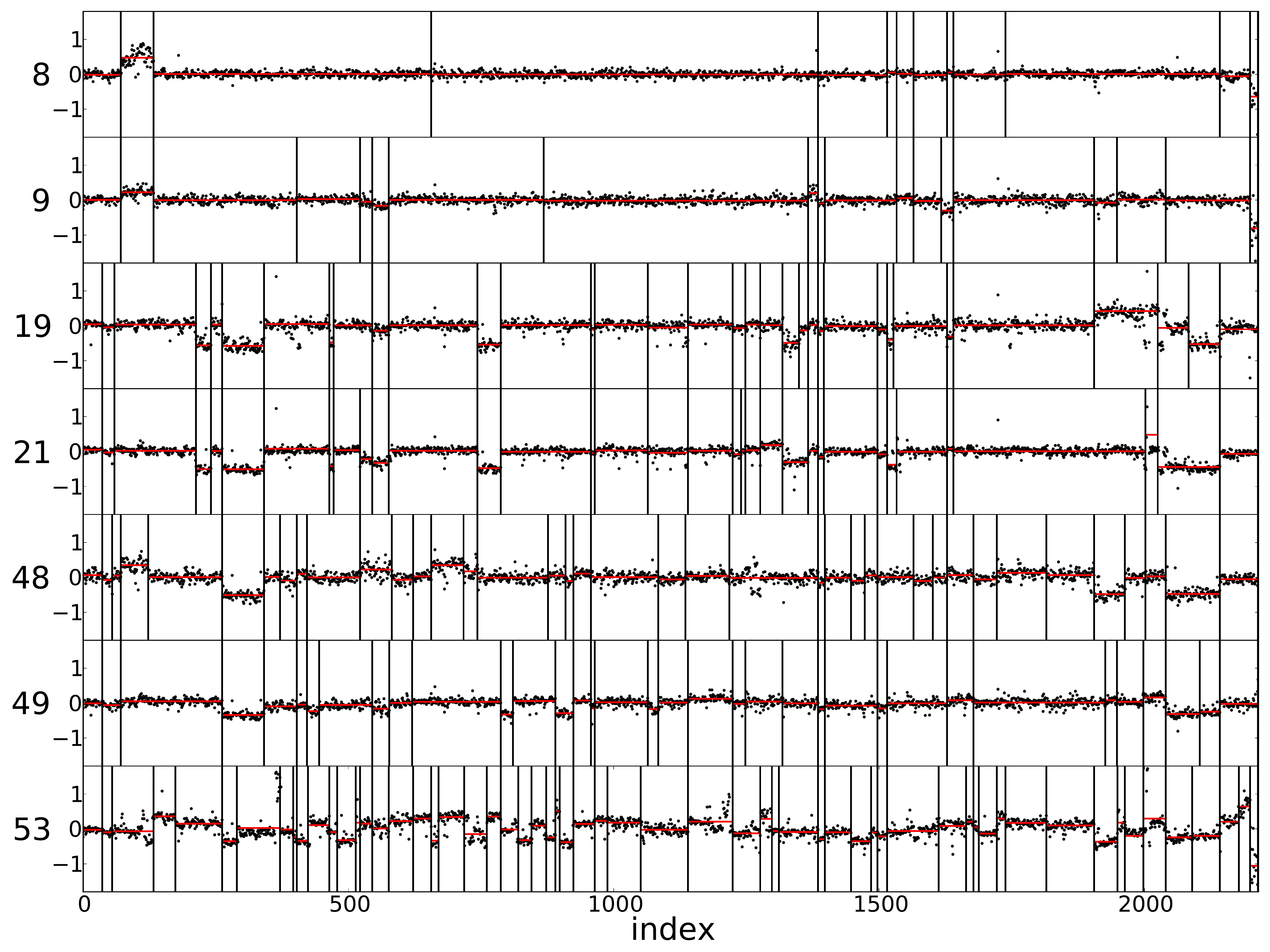}
\caption{Change-point detection on aCGH data, with the Bernoulli detector model for patients 8, 9, 21, 48, 49 and 53 jointly.}
\label{fig:aCGH_BD}
\end{center}
\end{figure}

In this application, we consider a publicly available dataset of Comparative
Genomic Hybridization array (aCGH) data, from bladder tumour samples,
presented in~\citep{Stransky2006}. The data come from the R package
ecp \citep{ecpR} and consists in 43 samples from tumours of different
patients, measuring the $\log_2$-ratio between the number of transcribed DNA
copies from tumorous cells and from a healthy reference. Each sample has 2215
probes. Changes in DNA copy number are one of the mechanisms responsible for
alterations in gene expression. Several mono or multivariate change-points
detection methods have already been applied to these data,
see~\citep{YutFong2011,Matteson2013,vert.2010.1}. The DNA sequences that are
over or under-expressed are jointly segmented for different patients. The
rationale behind the application of change-point detection algorithms to these
data
is to search for shared parts of DNA code, whose transcription is deregulated,
therefore likely to involve a similar mechanism of the tumour. Due to the complexity of the
algorithm, not all 43 samples are processed. One can notice on the
figure~\ref{fig:aCGH_BD} that some outliers exists in the data, but no
preprocessing is needed, thanks to the robustness to outliers of our proposed Bernoulli
Detector model.

\begin{figure}[htb]
\begin{center}
\includegraphics[width=1.0\linewidth]{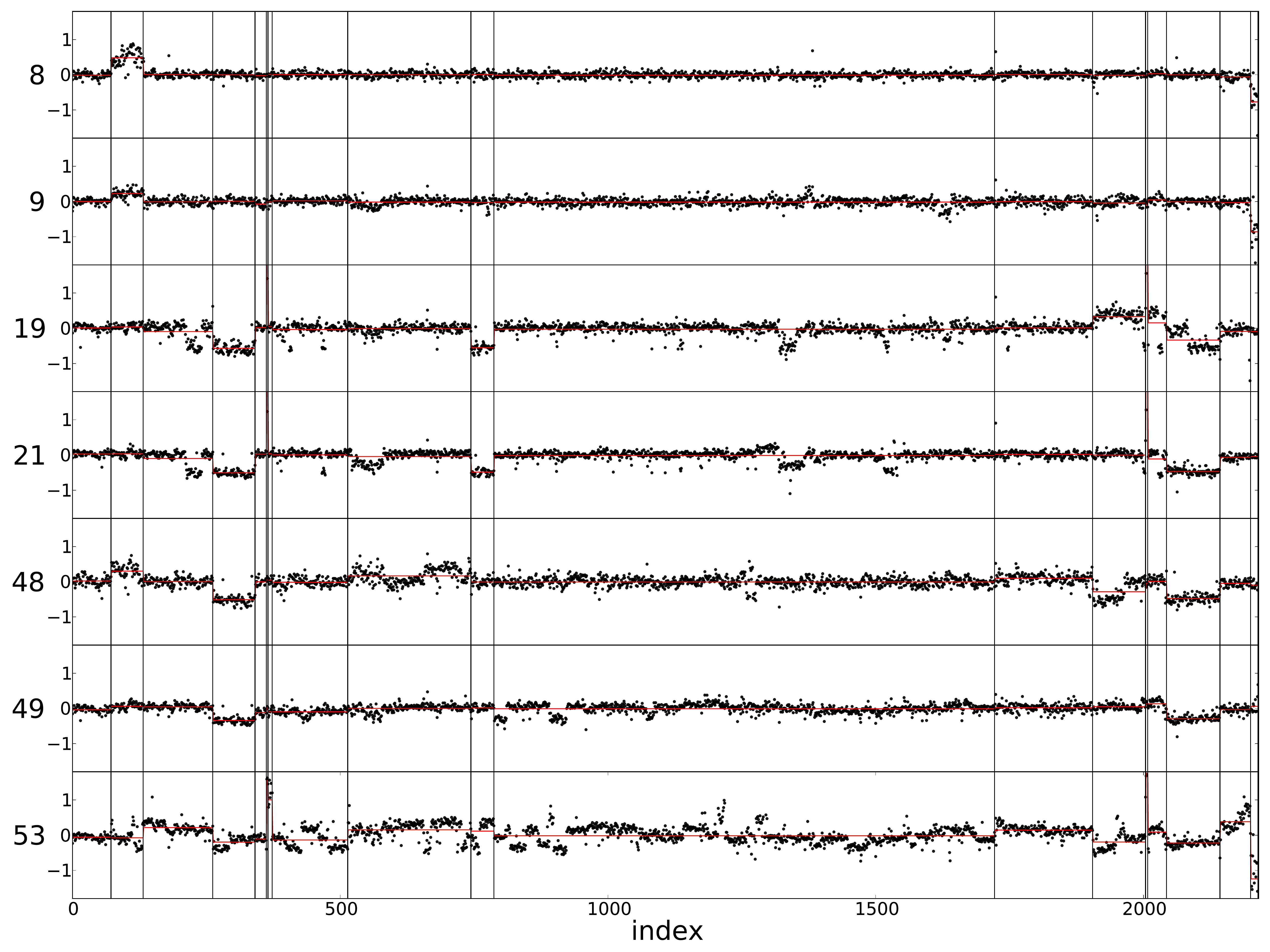}
\caption{Change-point detection on aCGH data, with the group fused lasso for patients 8, 9, 21, 48, 49 and 53 jointly.}
\label{fig:aCGH_GFL}
\end{center}
\end{figure}

To illustrate segmentations performed by counterparts change-point detection
algorithms, these data
are also processed by the group fused lasso method, presented
in~\citep{Bleakley2011}, which is an extension of the fused lasso to the
multivariate case. Please note that the purpose is to approximate all
time series by a unique piecewise-constant function, so all signals have the
same segmentation, even when some change-points are not shared by all signals,
but the accuracy of this method increases with the number of time series. We
used the MATLAB package GFLseg \citep{Bleakley2011} on the data to generate
the figure~\ref{fig:aCGH_GFL} on the same individuals. The change-points are
defined from the segments of the smoothed data. In the case of the Bernoulli
detector, the delimitations of each chromosome are not introduced. 

\begin{figure}[htb]
\begin{center}
\includegraphics[width=1.0\linewidth]{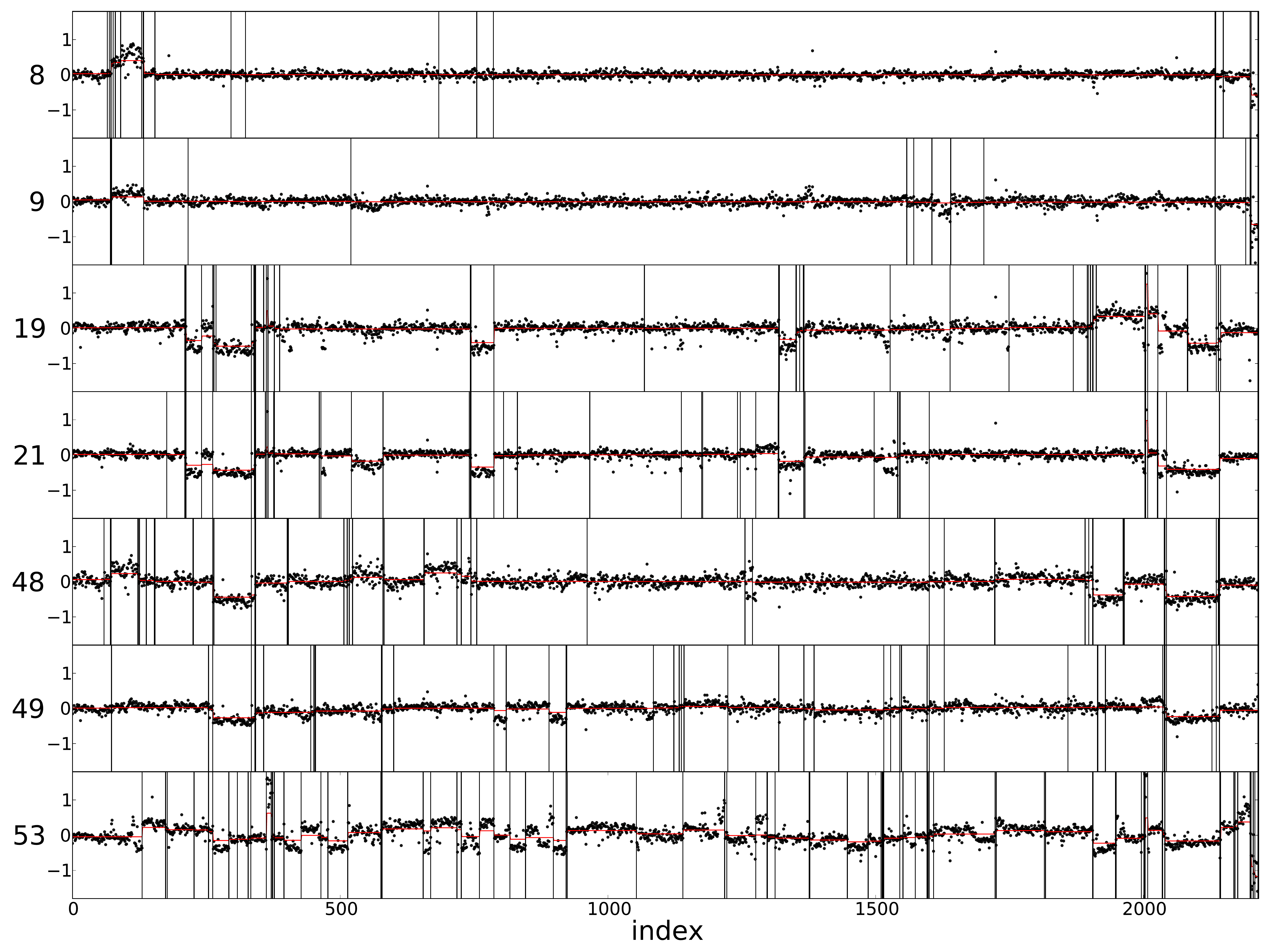}
\caption{Change-point detection on aCGH data, with the fused lasso on each patient 8, 9, 21, 48, 49 and 53, taken independently.}
\label{fig:aCGH_FL}
\end{center}
\end{figure}

The fused lasso \citep{Tibshirani2011} is also applied on each patient's data,
in a univariate framework, with the parameter $\lambda = 3.0$. The
change-points are detected when the difference between two successive values
of the approximate piecewise-constant function is over $10^{-5}$. The
resulting segmentation is shown in the figure~\ref{fig:aCGH_FL}.

The comparison between the solutions of the Bernoulli detector, the group
fused lasso and the fused lasso reveals three different objectives in
the change-point detection. With an independent univariate treatment, by the
fused lasso, each signal is taken individually, and the segmentation can be
optimized to fit the data at best. With the group fused lasso, the goal is to
identify some recurrent change-points positions, shared by several signals. It
leads to a unique segmentation, that is improved by an increasing number of
time series, but the less frequent features of some signals may be ignored.
Our approach is between these two: the parameter $\bP$ is learned to express
the probability that some signals share a change-point, and the parameter
$\bR$ gives the resulting segmentation.
\\

\setcounter{chapter}{4}
\refstepcounter{chapter}
\setcounter{equation}{4}
\noindent {\bf 5. Discussion and conclusion}
\label{chap:disc_concl}

The Bernoulli detector model has been introduced in this paper. It combines a
robust non-parametric statistical test and a Bayesian framework. Unlike
classical approaches described in the literature, only weak assumptions are made on
the change-points structure across multivariate time series. The proposed
method rather yields an estimate 
$\bP$, which describes the probability that two or more time series have
simultaneous events or not. Additionally the proposed approach
leads to two major advantages: first the composite marginal likelihood is
built from the $p$-values of a robust statistical test based on ranks. The
dependencies between the $p$-values are not taken into account, but the
inference function chosen, based on a Beta distribution of the $p$-values
under $H_1$ is empirically validated. Moreover the change-point model allows a
kind of control on the change-point detection by the acceptance level
$\alpha$. The impact of $\alpha$ has been measured on the FDR. However this
control has only be formalized in the case of a single change-point, due to
the difficulty to express the dependencies between
the $p$-values.

For the experiments, we opted for the Wilcoxon rank-sum test, that is
outlier-insensitive, and for which no normality assumption on data are
necessary. The method has been compared to other classical methods in the
univariate case, for a signal change-point. Their performances are similar for
normal data, and  the Bernoulli detector model is still efficient in term of
precision for data with outliers, whereas the fused
lasso~\citep{Tibshirani2011} and the Bernoulli Gaussian methods detect
non-existing change-points.

The second advantage is the introduction of the prior on the probability of
shared change-points on some signals~\citep{DobigeonApril}. This term is
learned by the model, and leads to information about an underlying dependency
structure between the time series. More complex data structures can be
analysed than the usual ones, where all signals are independent, then a
univariate method is applied, or fully connected, and then a unique
segmentation is search. Another way to use the Bernoulli detector model is to
introduce an informative prior on the dependency structure, and to remove the
configurations that are not possible or that have a small probability. Then the computation is much faster and
the segmentation of the time series is more precise. However it is not
conceivable to process jointly a large amount of time series.

A MCMC method has been chosen for the algorithm of the Bernoulli detector. The
use of this algorithm is then much longer than more classical ones, like the
fused lasso. In particular in the multivariate case, the complexity is linear
with the number of configurations tested. This impact can be reduced by the
choice of an in formative prior on the dependency structure when it is
possible. Another solution is the use of an approximate Gibbs sampler, that
reduces the number of operations. This approximation has been empirically
validated.

The Bernoulli detector has been applied on real datasets: some measurements of
electrical power consumption and some aCGH profiles data. With the first
application, we show the ability to learn the probability of simultaneous
change-points in particular groups of time series. With the second
application, we show an example of complex dependency structure, where each
signal has shared change-points with other signals, but has unique
change-points too. This approach differs from the usual ones
\citep{Tibshirani2011,Bleakley2011}. A clinical analysis of such results might
lead to information about common or unique deregulation mechanisms of the
genes expressions between several patients. From the estimated probabilities
$\bP$, it might be possible in future work to discuss about the estimation of
the causality between the time series, and to establish a representation of
the relationships between the signal with a graph.

Finally, we have introduced an innovative way to detect and localize multiple
change-points in multivariate time series, without assumption on the number of
change-points, and to learn alternatively the underlying dependency structure
between the signals. Despite the relatively long computation time, it presents
some noticeable advantages over other methods with interesting detection
performances. This approach may be complementary to other methods like the
fused of group lasso, especially in the case of multivariate data.
\\


\setcounter{chapter}{5}
\refstepcounter{chapter}
\noindent {\large\bf Acknowledgment}
GIPSA-lab is partner of the LabEx PERSYVAL-Lab (ANR-11-LABX-0025-01) funded by
the French program Investissement d'avenir.
\\
\par


\setcounter{chapter}{6}
\refstepcounter{chapter}
\noindent{\large\bf References}
\bibliographystyle{statsinica}
\renewcommand{\bibname}{}
\bibliography{references}


\vskip .65cm
\noindent
CEA, LIST, LADIS, 91191 Gif-sur-Yvette CEDEX, France
\vskip 2pt
\noindent
E-mail: flore.harle@cea.fr
\vskip 2pt

\noindent
Univ. Grenoble Alpes, GIPSA-Lab, F-38000 Grenoble, France
\vskip 2pt
\noindent
CNRS, GIPSA-Lab, F-38000 Grenoble, France
\vskip 2pt
\noindent
E-mail: florent.chatelain@gipsa-lab.grenoble-inp.fr
\vskip 2pt

\noindent
CEA, LIST, LADIS, 91191 Gif-sur-Yvette CEDEX, France
\vskip 2pt
\noindent
E-mail: cedric.gouy-pailler@cea.fr
\vskip 2pt

\noindent
Univ. Grenoble Alpes, GIPSA-Lab, F-38000 Grenoble, France
\vskip 2pt
\noindent
CNRS, GIPSA-Lab, F-38000 Grenoble, France
\vskip 2pt
\noindent
E-mail: sophie.achard@gipsa-lab.inpg.fr
\vskip .3cm


\end{document}